\documentclass[twocolumn,aps,showpacs,prx,superscriptaddress,preprintnumbers,amsmath,amssymb]{revtex4-2}
\pdfoutput=1
\usepackage{graphicx}
\usepackage{dcolumn}
\usepackage{bm}
\usepackage[shortlabels]{enumitem}
\usepackage[usenames,dvipsnames]{xcolor}
\usepackage{changes}

\makeatletter
\renewcommand\frontmatter@abstractwidth{\dimexpr\textwidth-1.2in\relax}
\makeatother

\definecolor{fuchsia}{rgb}{1.0, 0.0, 1.0}

\usepackage{subfigure}
\usepackage{enumitem}
\usepackage{bbm}

\usepackage[
  bookmarks=true,
  colorlinks,
  linkcolor=blue,
  urlcolor=blue,
  citecolor=blue,
  plainpages=false,
  pdfpagelabels,
  final,
  breaklinks=true
]{hyperref}
\hypersetup{
pdftitle={Avoiding local minima in Variational Quantum Algorithms with Neural Networks}, 
pdfauthor={Javier Rivera Dean}
}

\usepackage{titlesec}

\usepackage{array}

\usepackage{physics}

\begin{document}

\title{Avoiding local minima in Variational Quantum Algorithms with Neural Networks}

\author{Javier Rivera-Dean}
\email[\textbf{mail to: }]{javier.rivera@icfo.eu}
\affiliation{ICFO -- Institut de Ciencies Fotoniques, The Barcelona Institute of Science and Technology, 08860 Castelldefels (Barcelona), Spain}

\author{Patrick Huembeli}
\affiliation{Institute of Physics, \'Ecole Polytechnique F\'ed\'erale de Lausanne (EPFL), CH-1015 Lausanne, Switzerland}

\author{Antonio Acín}
\affiliation{ICFO -- Institut de Ciencies Fotoniques, The Barcelona Institute of Science and Technology, 08860 Castelldefels (Barcelona), Spain}
\affiliation{ICREA -- Institució Catalana de Recerca i Estudis Avançats, 08011 Barcelona, Spain}

\author{Joseph Bowles}
\affiliation{ICFO -- Institut de Ciencies Fotoniques, The Barcelona Institute of Science and Technology, 08860 Castelldefels (Barcelona), Spain}
\date{\today}

\begin{abstract}
Variational Quantum Algorithms have emerged as a leading paradigm for near-term quantum computation. In such algorithms, a parameterized quantum circuit is controlled via a classical optimization method that seeks to minimize a problem-dependent cost function. Although such algorithms are powerful in principle, the non-convexity of the associated cost landscapes and the prevalence of local minima means that local optimization methods such as gradient descent typically fail to reach good solutions. In this work we suggest a method to improve gradient-based approaches to variational quantum circuit optimization, which involves coupling the output of the quantum circuit to a classical neural network. The effect of this neural network is to peturb the cost landscape as a function of its parameters, so that local minima can be escaped or avoided via a modification to the cost landscape itself. We present two algorithms within this framework and numerically benchmark them on small instances of the Max-Cut optimization problem. We show that the method is able to reach deeper minima and lower cost values than standard gradient descent based approaches. Moreover, our algorithms require essentially the same number of quantum circuit evaluations per optimization step as the standard approach since, unlike the gradient with respect to the circuit, the neural network updates can be estimated in parallel via the backpropagation method. More generally, our approach suggests that relaxing the cost landscape is a fruitful path to improving near-term quantum computing algorithms. 
\end{abstract}

\maketitle

\section{INTRODUCTION}
The quantum computing community has now entered the so-called Noisy Intermediate-Scale Quantum (NISQ) computing era \cite{Preskill2018}, having access to programmable quantum devices which have on the order of hundreds of non-error-corrected qubits. Although scientists have shown strong evidence of quantum computational supremacy \cite{Google2019,zhong2020quantum} for some class of problems within this paradigm, most of the well-known quantum algorithms \cite{Montanaro2016} for problems of practical interest have necessities that go well beyond the capabilities of NISQ devices. A huge effort has therefore been made in recent years to develop algorithms that can suit the capabilities of NISQ devices

The majority of such NISQ-ready algorithms fall into the class of \emph{variational quantum algorithms} \cite{Peruzzo2014, McClean2016, Endo2020, Bharti2020(1), Bharti2020(2)}. In this approach, one defines a problem-dependent cost function that is evaluated on the output of a classically parameterized quantum circuit. These parameters are then updated iteratively via a classical optimization method that interfaces with the quantum circuit and seeks to minimize the cost. Although the method is expected to be powerful in principle, one typically faces the problem of non-convexity in the cost function landscape, whereby local optimization methods such as gradient descent, commonly encounter sub-optimal local minima. If such local optimization methods are to succeed, it is therefore crucial to develop techniques that tackle the problem of local minima in the optimization landscape. 

In this work we focus on gradient based methods for quantum circuit optimisation, due to their effectiveness in solving optimization problems for large parameter spaces, their widespread use in variational quantum algorithms, and their computational efficiency and clear convergence criteria \cite{Ruder2016}. The majority of gradient based approaches keep the cost landscape fixed and incorporate elements such as momentum, adaptive step sizes, and randomization to overcome or escape local minima. In this work we introduce a conceptually different method that involves modifying the cost landscape itself in a controlled manner. This is achieved by coupling a classical neural network to the output of the quantum circuit, which can continuously deform the cost landscape as a function of its weights by postprocessing the measurements outcomes obtained at the end of the quantum circuit. The addition of the neural network weights to the problem results in a higher level of parameterization which---as is known to be the case in classical \cite{overparam} and quantum machine learning \cite{overparam2}---can be expected to help in avoiding local minima. It is this intuition that is the driving force behind our work.

In our approach the neural networks weights are updated by applying the backpropagation method on batches of bit strings that are sampled from the quantum circuit and processed by the neural network. Since backpropagation allows for gradient components to be calculated in parallel, this means that for large problems, the gradient with respect to the neural network parameters can be estimated much more efficiently than the gradient with respect to the quantum circuit parameters, which requires sampling two different quantum circuit for each parameter \cite{Schuld2019}. This means that our method requires essentially the same number of quantum circuit evaluations as the \emph{standard approach} per optimization step. Here by \emph{standard approach} we refer to the simplest gradient-based optimisation procedure of parameterized quantum circuits without additional post-processing: i.e.\ where the gradient with respect to the quantum circuit parameters is estimated and used to step the circuit parameters, and all parameters are initialized randomly.

We define two algorithms that take advantage of our architecture. The first acts as a means to escape a local minimum in which a standard variational quantum algorithm has become stuck. The second iteratively optimizes the cost landscape throughout the entire optimization process as a means of evading local minima. 

We benchmark our algorithms on instances of the Max-Cut optimization problem \cite{Zhou2020} via numerical simulations, considering two classes of circuit ans\"atzen. Interestingly, we show that for a simple problem, the cost landscape is modified in such a way that connects local minima of the original landscape to the global minimum of the problem, thus leading to much better solutions. For more complex problems, we show that our algorithms improve over the standard approach by finding both deeper minima and lower cost values on average. 

The article is structured as follows: in Sec.~\ref{Methods} we briefly review the basics of Variational Quantum Algorithms and Artificial Neural Networks, and then outline our approach and algorithms. In Sec.~\ref{Results} we present the results of our benchmarks and the effect of hyperparameter selection, and discuss possible lines for further research in Sec.~\ref{discussion}. 

\section{METHODS}\label{Methods}

Our architecture combines variational quantum algorithms and artificial neural networks. In this section we present these two ingredients separately and then discuss how to use them together to solve optimization problems in NISQ devices.

\subsection{Variational quantum circuit optimization}\label{VQAs}
The variational method \cite{Sakurai, Griffiths} is one of the most widely used techniques for finding the ground state of a system with a Hamiltonian $H$. The method defines a parameter-dependent quantum state, $\ket{\psi(\boldsymbol{\theta})}$, and seeks the optimal $\boldsymbol{\theta^*}$ that minimizes the energy of $H$:
\begin{equation}\label{Variational:Method}
    C(\boldsymbol{\theta^*}) = \min_{\boldsymbol{\theta}} C(\boldsymbol{\theta}) =
       \min_{\boldsymbol{\theta}} \mel{\psi(\boldsymbol{\theta})}{H}{\psi(\boldsymbol{\theta})}.
\end{equation}
Here $\boldsymbol{\theta} = (\theta_1, \dots, \theta_P)$ is a list of $P$ real parameters and the function $ C(\boldsymbol{\theta})$ will be referred to as the cost function. From the Rayleigh-Ritz variational principle, a lower bound to Eq.~\eqref{Variational:Method} is given by the ground state energy of the system, as the ground state may not be written in terms of the parameter-dependent state $\ket{\psi(\boldsymbol{\theta})}$.

Variational Quantum Algorithms (VQAs) \cite{Cerezo2020} attempt to solve the optimization problem of Eq.~\eqref{Variational:Method} using a quantum-classical hybrid approach. They prepare and measure states $\ket{\psi(\boldsymbol{\theta})}$ via a parametrized quantum circuit, usually referred to as a variational quantum circuit (VQC). 
This circuit is described by a unitary transformation $U(\boldsymbol{\theta})$ that acts on an easy-to-prepare initial state $\ket{\psi_0}$, where $U(\boldsymbol{\theta})$ is typically composed as a sequence of unitaries from a predefined gate set that each depend on a single parameter $\theta_i$:
\begin{equation}
    \ket{\psi(\boldsymbol{\theta})}
        = U(\boldsymbol{\theta}) \ket{\psi_0}
        = \prod_{i=1}^P U_i(\theta_i) \ket{\psi_0},
\end{equation} 
with $P$ the total number of parameters. In order to find good parameters that minimize the cost, the system is interfaced with a classical optimization method, which controls the preparation of states $\ket{\psi(\boldsymbol{\theta})}$ and iteratively updates a set of candidate solution parameters for Eq.~\eqref{Variational:Method}, see Fig.~\ref{Fig1}~(a).

There exists a large variety of classical optimization methods that could be used in a VQA. In this work, we consider gradient based methods \cite{Kelley1960, Ruder2016}, which are currently the most common method in VQAs, although derivative-free methods \cite{derivativefree,McClean2016,Nannicini2019} can also be applied. The standard first order gradient descent method involves stepping the circuit parameters in the opposite direction of the gradient of the cost:
\begin{equation}\label{Grad:method}
    \theta_j^{(i+1)} =
        \theta_j^{(i)} - \eta \pdv{C(\boldsymbol{\theta})}{\theta_j},
\end{equation}
where $\theta_j^{(i)}$ is the value of parameter $\theta_j$ at the $i$th iteration of the algorithm, and $\eta$ is a predefined step size. One is thus required to estimate the partial derivatives of the cost function with respect to the circuit parameters. The best known method that achieves this for a wide range of circuit architectures and is suitable to near-term devices is the parameter-shift rule \cite{Schuld2019, Mitarai2018}. Here one considers gates of the form $U(\theta_j) = e^{-i\theta_j O}$, where $O$ is an Hermitian operator with at most two unique eigenvalues $\pm r$. This requirement involves quantum circuit architectures with two qubit gates like the QAOA architecture \cite{Sweke2021}. The derivative $\pdv*{C(\boldsymbol{\theta})}{\theta_j}$ is then found to be equal to
\begin{equation}\label{Param:shift}
    \begin{aligned}
    \pdv{C(\boldsymbol{\theta})}{\theta_j}
        =\ & r\big[
                \bra{\psi_0}
                U^\dagger(\boldsymbol{\theta}+\boldsymbol{e}_j) H
                U(\boldsymbol{\theta}+\boldsymbol{e}_j)\ket{\psi_0}\\
                &-
                \bra{\psi_0}U^\dagger(\boldsymbol{\theta}-\boldsymbol{e}_j) H
                U(\boldsymbol{\theta}-\boldsymbol{e}_j)\ket{\psi_0}\big],
    \end{aligned}
\end{equation}
where $\boldsymbol{\theta}\pm\boldsymbol{e}_j=(\theta_1,\cdots,\theta_j\pm\frac{\pi}{4r},\cdots,\theta_P)$. Thus, the gradient with respect to each parameter is given by the difference between the cost of two \emph{parameter-shifted} circuits. 
Note that the cost must be estimated by sampling the quantum circuit $n$ times, where $n$ defines the precision of the estimation. Since each partial derivative requires shifting the parameters differently, the number of circuit evaluations needed to estimate the full gradient to this precision is $2nP$. We note quantum backpropagation algorithms that allow for parallel gradient estimation have been introduced \cite{Verdon2018}, although are less suitable for near term devices since the optimization parameters need to be stored coherently in additional ancilla systems. 

Here we have discussed the case of parameter shift rule since it is very well suited for the quantum circuit architectures we consider in this manuscript. Yet, and as mentioned above, it is only valid for those situations where the involved quantum gates have no more than two spectral levels. For more complicated situations other techniques should be considered, like the ones presented in \cite{Williams2011}. However, we would like to remark that this does not affect the algorithms developed in this article since, as we will see now, the neural network optimization can be done separately from the one of the variational quantum circuit.

\subsection{Artificial neural networks}\label{QAOA}
Artificial Neural Networks are a computational model inspired by the architecture of biological neural connections, where each neuron is connected to other multiple neurons and transmits information in the form of electric pulses. They have been applied successfully to many different problems such as computer vision \cite{Ciregan2012}, reinforcement  learning \cite{Vinyals2019} and natural language processing \cite{Brown2020}. Furthermore, they have been employed to study phenomena in many-body physics, being useful for the study of phase transitions \cite{Carrasquilla2017}, ground state energy minimization \cite{Carleo2017}, and quantum circuit simulation \cite{Carrasquilla2019}. With respect to VQAs, neural networks have been used for meta-learning the optimization method \cite{Wilson2019,Verdon2019} and searching for solutions to combinatorial optimization problems \cite{Wauters2020}. 

\begin{figure}
	\includegraphics[width=\columnwidth]{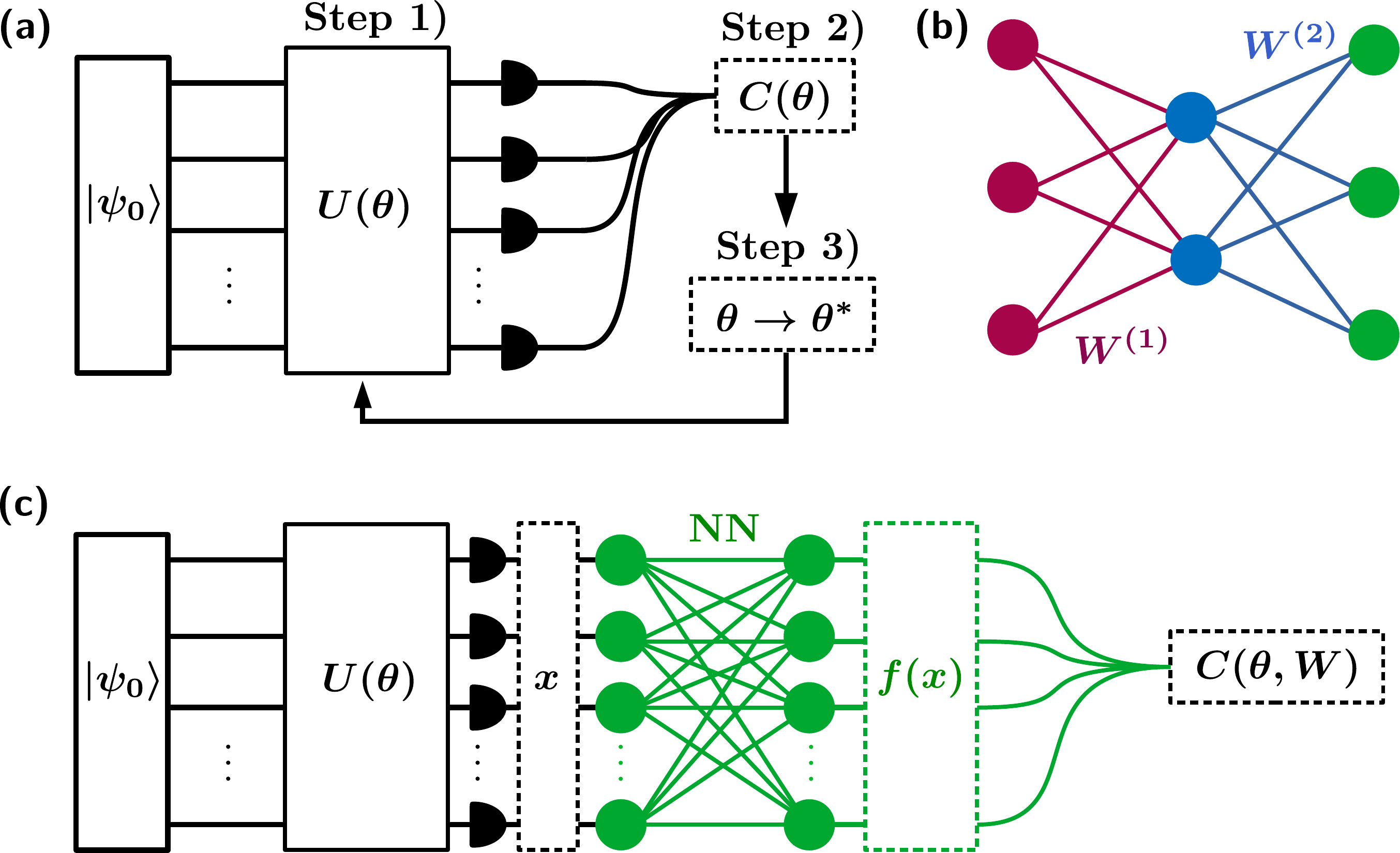}
	\caption{(a) General structure of a VQA. A VQC prepares and measures a multi-qubit state via a parameterized unitary, from which a cost function is evaluated. A classical optimization loop is interfaced with the VQC which updates the parameters in order to minimize the cost. (b) A three layer feedforward neural network consisting of an input layer (in purple), a single hidden layer (in blue) and an output layer (in green). (c) Architecture used for our algorithms. We initialize the system with a VQA similar to the one shown in (a), but now its outputs are post-processed by a feedforward neural network (in green), composed of a fully connected input and output layer, which adds an extra set of parameters described by the matrix $W$ that enter in our cost function.} 
	\label{Fig1}
\end{figure}

A \emph{feedforward neural network} is a simple and widely used neural network architecture. Such a neural network returns an output vector $\boldsymbol{y}$ that is the result of a sequence of matrix multiplications and element-wise nonlinear transformations, called activation functions, applied to an input vector $\boldsymbol{x}$:
\begin{align}
    \boldsymbol{y}=a^{(L)}(W^{(L)}(\cdots a^{(2)}(W^{(2)}(a^{(1)}(W^{(1)}\boldsymbol{x})))\cdots)),
\end{align}
see Fig.~\ref{Fig1}~(b) for a schematic representation. Here the $W^{(i)}$ denote the matrices and the $a^{(i)}(\cdot)$ the activation functions.  Common choices of activation function are the Heaviside function (resulting in a \emph{perceptron} \cite{Rosenblatt1958, NNielsen} neural network), the sigmoid function or the linear rectifier function (ReLU). 
In some models one also adds an additional bias vector after each matrix multiplication, but we do not consider this in our work. From hereon we focus on the simplest model of a single layer feedforward neural network, i.e.\ the output is given by 
\begin{align}
    \boldsymbol{y}(\boldsymbol{x})=a(W(\boldsymbol{x}))\equiv f_W(\boldsymbol{x})
\end{align}
and we refer to the elements of the matrix $W$ as \emph{weights}.

The output $\boldsymbol{y}$ is used to evaluate a cost function $C(\boldsymbol{y})=C(f_W(\boldsymbol{x}))$, where $C(\boldsymbol{y}):\mathbbm{R}^N \to \mathbbm{R}$. In this work we are interested in a scenario in which $\boldsymbol{x}$ is not a fixed vector but a classical random variable $\boldsymbol{X}\sim p(\boldsymbol{x})$ over a discrete set of vectors $\{\boldsymbol{x}\}$. Sampling this $\boldsymbol{x}$ and passing it through the neural network results in a new random variable $f_W(\boldsymbol{X})$ with an associated mean cost
\begin{align}\label{stochcost}
    \mathcal{C}(W)=\sum_{\boldsymbol{x}}p(\boldsymbol{x})C(f_W(\boldsymbol{x}))
\end{align}
The above cost can be minimized via a gradient descent procedure that iteratively updates the weights. The gradient vector of the above with respect to the weights is 
\begin{align}\label{stochgrad}
    \nabla_W \mathcal{C}(W)=\sum_{\boldsymbol{x}}p(\boldsymbol{x})\nabla_W C(f_W(\boldsymbol{x})).
\end{align}
For fixed $\boldsymbol{x}$ the gradient vector $\nabla_W C(f_W(\boldsymbol{x}))$ can be calculated at a similar computational cost to that of evaluating $C(f_W(\boldsymbol{x}))$ via the celebrated backpropagation algorithm \cite{Rumelhart1986}. Thus, from Eq.~\eqref{stochgrad}, the gradient of $\mathcal{C}$ can be estimated by sampling a batch of vectors $\boldsymbol{x}$ and summing the corresponding gradient vectors for each $\boldsymbol{x}$. This gradient can then be used in a gradient descent procedure to minimize the cost in Eq.~\eqref{stochcost}. 

\subsection{Neural network assisted variational quantum optimization}\label{Hybrid}
We now present our neural network assisted VQA algorithms. We consider a variational problem as in Eq.~\eqref{Variational:Method}
\begin{equation}\label{diag_problem}
       \min_{\boldsymbol{\theta}} \mel{\psi(\boldsymbol{\theta})}{H_D}{\psi(\boldsymbol{\theta})},
\end{equation}
where $H_D$ is diagonal in the computational basis:
\begin{align}\label{costH}
    H_D =\sum_{\boldsymbol{x}\in\{\pm 1\}^N}
    \! \! \! \! C(\boldsymbol{x})\ket{\boldsymbol{x}}\bra{\boldsymbol{x}}.
\end{align}

An extension of the algorithm to the case where $H$ is not diagonal in the computational basis but given by a sum of local Hamiltonians is straightforward and is discussed in Sec.~\ref{discussion}. Mappings of classical optimization and classical machine learning problems to VQAs are typically expressed in the form of Eq.~\eqref{diag_problem}. As mentioned, many works have studied finding solutions to such problems using a standard VQA architecture. Here we describe how to combine such architectures with classical feedforward neural networks. 

The algorithms we present are based on the architecture shown in Fig.~\ref{Fig1}~(c). First, a state $\ket{\psi(\boldsymbol{\theta})}$ is prepared by a generic VQC and an $N$ qubit computational basis measurement is made, producing the vector $\boldsymbol{x}\in\{-1,+1\}^N$. This vector is fed into a single layer feedforward neural network, leading to the output vector $f_W(\boldsymbol{x})=\tanh(W\boldsymbol{x})\in\mathbb{R}^N$. Here we have used the hyperbolic tangent function as an activation function, which serves as a continuous relaxation to the sign function. Finally, the cost $C$ is evaluated on the output vector. Sampling from the quantum circuit, one obtains the mean cost
\begin{align}\label{costfull}
    \mathcal{C}(\boldsymbol{\theta},\boldsymbol{W})=\sum_{\boldsymbol{x}\in\{\pm 1\}}p(\boldsymbol{x}\vert\boldsymbol{\theta})C(f_W(\boldsymbol{x}))
\end{align}
where the probabilities $p(\boldsymbol{x}\vert\boldsymbol{\theta})$ are given from the VQC as 
\begin{equation}
     p(\boldsymbol{x}\vert\boldsymbol{\theta})
    = \vert \! \braket{\psi(\boldsymbol{\theta})}{\boldsymbol{x}} \! \vert^2.
\end{equation}
In this context, the quantum circuit can be seen as a $\boldsymbol{\theta}$-dependent \emph{sample generator} for the neural network, and the mean cost takes the same form as in Eq.~\eqref{stochcost}. Note that Eq.~\eqref{costfull} can also be written 
\begin{equation}\label{Cost:fun:Hyb}
     \mathcal{C}(\boldsymbol{\theta}, W)
    = \mel{\psi(\boldsymbol{\theta})}{H(W)}{\psi(\boldsymbol{\theta})},
\end{equation}
with 
\begin{equation}\label{H(W):def}
    H(W) = 
        \sum_{\boldsymbol{x}\in\{\pm 1\}} C\big(f_W(\boldsymbol{x})\big) \dyad{\boldsymbol{x}}.
\end{equation}

Thus, one can see that changing the weights $W$ modifies the energy landscape of the cost with respect to the quantum circuit parameters. Eqs.~\eqref{costfull} and \eqref{Cost:fun:Hyb} also imply that the gradients $\nabla_{\boldsymbol{\theta}}\mathcal{C}(\boldsymbol{\theta}, W)$ and $\nabla_W \mathcal{C}(\boldsymbol{\theta}, W)$ can both be estimated by sampling the quantum circuit: from Eq.~\eqref{Cost:fun:Hyb} one can estimate $\nabla_{\boldsymbol{\theta}}\mathcal{C}(\boldsymbol{\theta}, W)$ through the parameter-shift rule of  Eq.~\eqref{Grad:method} with the Hamiltonian $H(W)$, and from Eq.~\eqref{Cost:fun:Hyb} one can estimate $\nabla_{W}\mathcal{C}(\boldsymbol{\theta}, W)$ by accumulating the gradient through the backpropagation rule of Eq.~\eqref{stochgrad}. 

At this point it is relevant to notice that the estimation of $\nabla_{W}\mathcal{C}(\boldsymbol{\theta}, W)$ is much more efficient in terms of the number of calls to the quantum circuit.
In particular, estimating $\nabla_{\boldsymbol{\theta}}\mathcal{C}(\boldsymbol{\theta}, W)$ requires sampling from two parameter-shifted circuits for each of the $P$ parameters, whereas estimating $\nabla_{W}\mathcal{C}(\boldsymbol{\theta}, W)$ involves sampling from a single quantum circuit with fixed parameters $\boldsymbol{\theta}$. Since $P$ generally grows with at least the number of qubits, the number of circuit calls needed to calculate $\nabla_W \mathcal{C}(\boldsymbol{\theta}, W)$ relative to $\nabla_{\boldsymbol{\theta}} \mathcal{C}(\boldsymbol{\theta}, W)$ becomes negligible for large circuits. This difference can be seen as a consequence of the fact that the backpropagation algorithm allows for partial derivatives to be calculated in parallel, which is not the case for the parameter-shift rule. Thus, if one fixes a precision in the estimation of the gradients with respect to both the circuit parameters and the neural network parameters, then the number of circuit evaluations needed for computing the gradient (e.g. via the parameter-shift rule) with respect to the circuit parameters scales with the total number of parameters. On the other hand, and due to the efficiency of the backpropagation algorithm, the number of circuit evaluations needed for computing the gradient with respect to the neural network is constant, meaning by this that it is independent of total number of quantum circuit parameters. Thus, we only need to sample from the variational quantum circuit in order to generate a batch of bitstrings with respect to which the neural network parameters are updated.

\begin{figure}
	\includegraphics[width=\columnwidth]{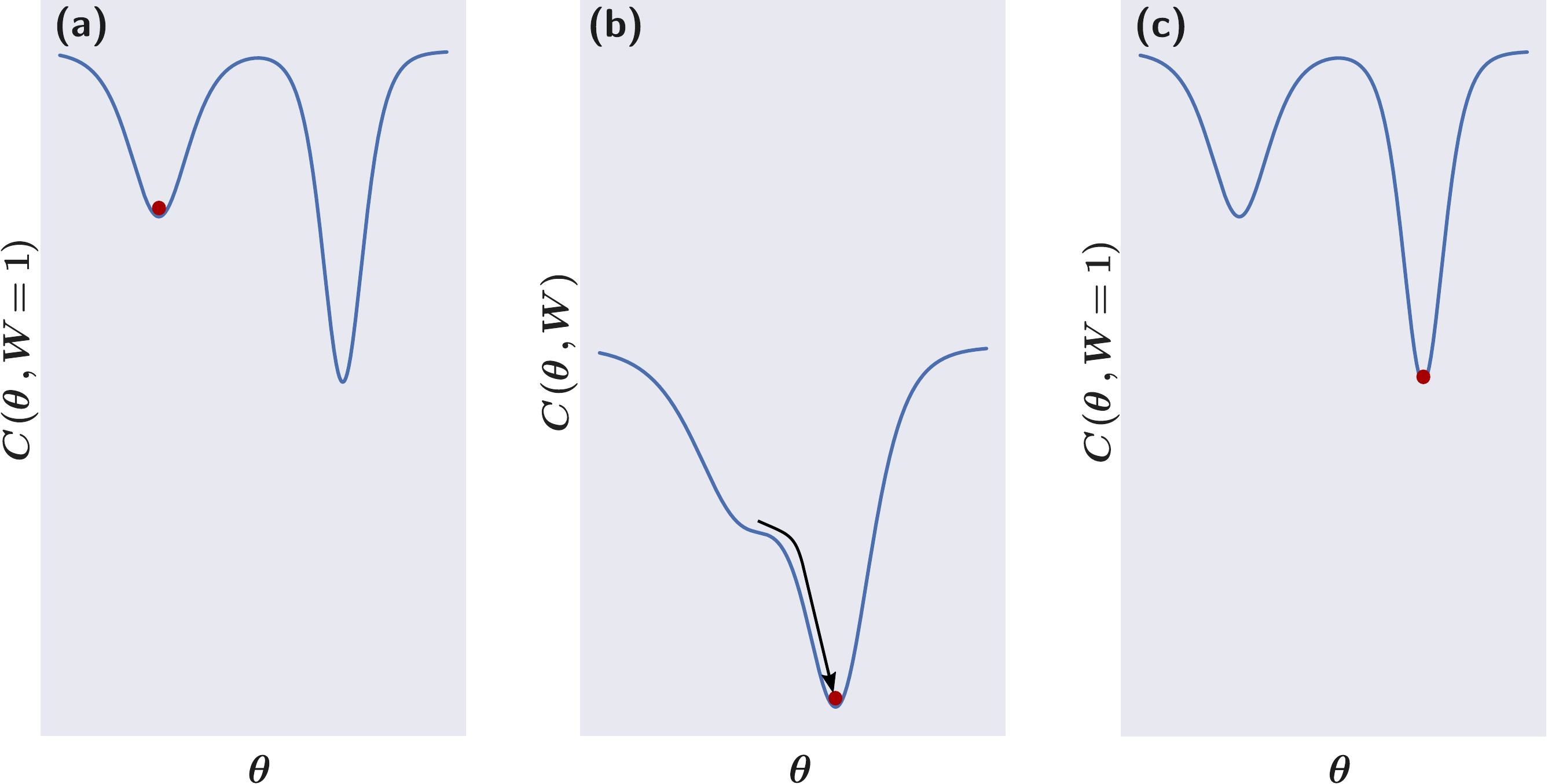}
	\caption{Representation illustrating the main idea of our algorithms. (a) The original cost landscape in which the VQC has become trapped in a local minimum. (b) By updating the neural network weights, one perturbs the cost landscape so that the VQC is no longer in a local minimum. (c) The VQC is optimized in the new landscape, potentially finding a better local minimum. } 
	\label{Fig2}
\end{figure}

As mentioned, from Eq.~\eqref{Cost:fun:Hyb} the weights $W$ modify the cost landscape with respect to the quantum circuit parameters. In particular, if one fixes $W=\mathbb{I}$ then the neural network is proportional to the identity transformation, and the cost is equivalent to the standard VQA cost, whereas for generic $W$ the landscape with respect to $\boldsymbol{\theta}$ is modified. This naturally opens the opportunity to design methods that improve the performance of VQAs by exploiting this ability, as shown in Fig.~\ref{Fig2}. Furthermore, since gradient estimations with respect to the neural network weights require relatively few circuit evaluations, such algorithms can be as computationally efficient as standard VQA algorithms. This leads us to the central question of our work:
\\[5pt]
\noindent{\centering{
\emph{Can neural networks help a VQA obtain significantly better solutions to Eq.~\eqref{diag_problem}, with a negligible increase in the number of calls to the quantum circuit?}}}
\\[5pt]
We present two algorithms that achieve this goal. In the first algorithm the neural network is employed as a means to escape local minima encountered by standard VQA algorithms, and is therefore only used when such minima are encountered. In the second algorithm, the neural network modifies the cost landscape during the entire optimization and thus guides the VQA optimization by continuously modifying the cost landscape. Since our aim is for the neural network to aid the VQA optimization and not solve the problem itself, we design our algorithms such that $W=\mathbb{I}$ upon termination. This ensures that the final parameters $\boldsymbol{\theta}$ encode a good solution in the original landscape.

\subsubsection{Algorithm 1: \texttt{ESCAPE} algorithm}
The purpose of this algorithm is to help the VQA escape from local minima. The basic idea is to run a standard VQA without help from the neural network until convergence to a (local) minimum (step 2 below). At that point, the VQC parameters $\boldsymbol{\theta}$ are held fixed, and the cost landscape is modified by engaging the neural network and optimizing its weights (step 3). In this new landscape the VQC will generally no longer be in a local minimum. 
The VQC parameters are then optimized again, during which the cost landscape is brought back to the original one by gradually removing the effect of the neural network (step 4). At this point, the VQC has hopefully escaped the local minimum that it first encountered, and so the circuit parameters are updated again in the original landscape, until another local minimum is reached. The precise steps composing the \texttt{ESCAPE} algorithm are:
\begin{enumerate}[1)]
    \item Initialize $\boldsymbol{\theta}$ and set $W=\mathbb{I}$.
    \item Update $\boldsymbol{\theta}$ via a gradient descent until convergence to a local minima.
    \item Update $W$ via a gradient descent procedure for $M$ steps and define $W_0$ as the weight matrix after the last step.
    \item For $t=1,\cdots,T$:
    \begin{enumerate}
        \item Set $W=(1-g(t))W_0 + g(t)\mathbbm{1}$, \\[3pt] where $g(0)=0$ and $g(1)=1$. 
        \item Perform a gradient descent update step of $\boldsymbol{\theta}$.
    \end{enumerate}
    \item Update $\boldsymbol{\theta}$ via gradient descent until convergence to a (potentially different) local minimum.
    \item Compare the energies obtained in step 2) and step 5) and take the lowest one.
\end{enumerate}

\begin{figure}
	\includegraphics[width=1\columnwidth]{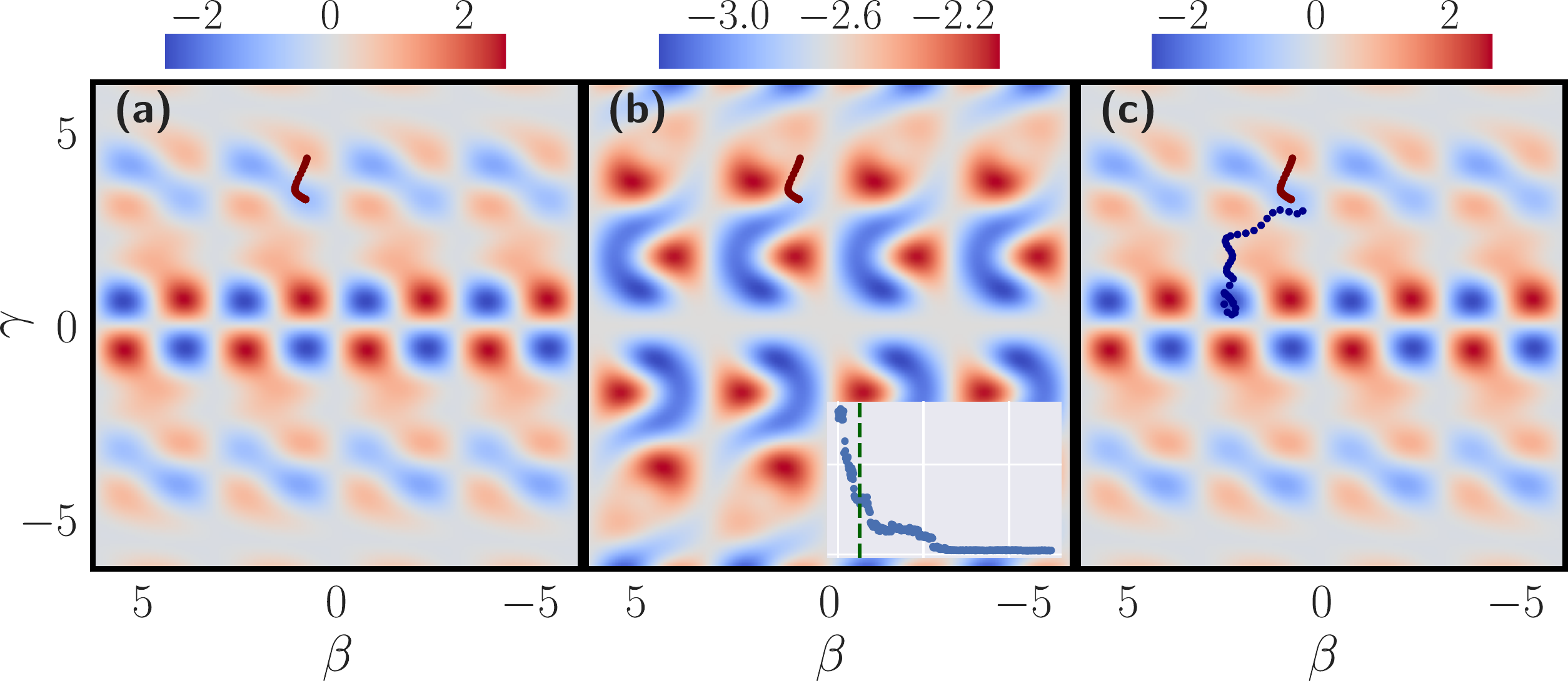}
	\caption{\texttt{ESCAPE} algorithm carried out with a QAOA architecture, with two optimization parameters $\gamma$ and $\beta$. (a) Original energy landscape before the neural network minimization. The dark red curve represents step 2), where a given quantum circuit initialization, originally placed close to a maximum, converges to a local minimum. (b) Obtained energy landscape after the neural network optimization (step 3). The small subplot shows the neural network optimization; the horizontal line represents the point at which training was stopped. (c) Convergence of the QAOA to a global minimum (dark blue line) during steps 4-5.}
	\label{Fig3}
\end{figure}

In this algorithm the neural network works as an \emph{assistant} of the VQC rather than the \emph{problem solver}, since it is only used to escape local minima in the original VQC cost landscape. This bears some similarity to standard basin hopping algorithms \cite{Olson2012}, a widely used method in non-convex optimization that attempts to escape local minima by randomly perturbing the parameters when stuck in a local minimum. An important difference between basin hopping and our method is that whereas the standard basin hopping algorithm randomly perturbs the parameters, we keep the parameters fixed and modify the cost landscape in a way that is dependent on both the local parameters and the problem. Such methods may therefore be better suited to successfully escaping minima since they are tuned to the problem and local information at hand. An example showing how this method works in practice is presented in Fig.~\ref{Fig3}. Remarkably, the effect of optimizing the neural network in step 3 is to connect the local minimum to the global minimum, which is then found in steps 4-5. The precise mechanism behind this phenomenon and whether it is generic is still unclear, however it may be related to the existence of minima pools \cite{Draxler2019} in neural network training landscapes. 

\subsubsection{Algorithm 2: \texttt{GUIDE} algorithm}
In our second algorithm, both the neural network weights and VQC parameters are updated in every optimization step, rather than only when $\boldsymbol{\theta}$ is placed at a (local) minimum. The cost landscape is therefore changed during the entire optimization process as a means to guide the VQC optimization in real-time. We do not however want the neural network to dramatically change the cost landscape, since our aim is still to arrive at a good solution in the unperturbed landscape at the end of the optimization. For this reason, we introduce a modified cost $\mathcal{C}_{\alpha}$:
\begin{align}
     \mathcal{C}_{\alpha}(\boldsymbol{\theta}, W)
        = \mathcal{C}(\boldsymbol{\theta}, W)
            + \alpha \vert W-\mathbb{I} \vert,
\end{align}
where $\alpha\geq 0$ and the term $\vert W-\mathbb{I} \vert = \sum_{ij} \vert w_{ij}-\delta_{ij}\vert$ acts as a regularization that penalizes the weight matrix being far from the identity, and therefore limits how much it can modify the cost landscape. Here, the parameter $\alpha$ is a hyperparameter that should be tuned to the specific problem. The precise steps of the \texttt{GUIDE} algorithm are as follows:
\begin{enumerate}[1)]
    \item Initialize $\boldsymbol{\theta}$ and set $W=\mathbb{I}$.
    \item Estimate $\nabla_{\boldsymbol{\theta}}\mathcal{C}_{\alpha}(\boldsymbol{\theta}, W)$ and perform a gradient update step of $\boldsymbol{\theta}$.
    \item Estimate $\nabla_{W}\mathcal{C}_{\alpha}(\boldsymbol{\theta}, W)$ and perform a gradient update step of $W$.
    \item Repeat steps 2 - 3 until convergence, and define $W_0$ as the weight matrix after the last update. 
    \item For $t=1,\cdots,T$:
    \begin{enumerate}
        \item Set $W=(1-g(t))W_0 + g(t)\mathbbm{1}$, \\[3pt] where $g(0)=0$ and $g(1)=1$. 
        \item Perform a gradient descent update step of $\boldsymbol{\theta}$.
    \end{enumerate}
    \item Update $\boldsymbol{\theta}$ via gradient descent until convergence to a (potentially different) local minimum.
\end{enumerate}
As noted previously, if the gradients $\nabla_{\boldsymbol{\theta}}\mathcal{C}_{\alpha}(\boldsymbol{\theta}, W)$ are calculated via the parameter-shift rule, the addition of the neural network updates (step 3) comes at essentially no extra cost. The \texttt{GUIDE} algorithm can therefore replace a standard VQA algorithm without any changes to the VQC architecture, and essentially the same number of quantum circuit calls per optimization step. 

\section{RESULTS}\label{Results}
\subsection{Performance of our algorithms compared to the standard approach}
We benchmark our algorithms against the standard gradient descent VQA algorithm with fixed cost landscape. For the gradient descent subroutine in both the standard and our algorithms, we consider both standard gradient descent (i.e. the update rule of Eq.~\eqref{Grad:method}), and the so-called Adam gradient descent method \cite{Kingma2017}, which features an adaptive step size and momentum. 
\subsubsection{Optimization problems}
The problems we study are instances of the Max-Cut optimization problem. Given a graph of $N$ nodes and a set $E$ of weighted edges, where we denote the weight of an edge $(i,j)$ by $J_{ij}$ with $(i,j) \in E$, the task is to find a partition of the nodes into two sets, such that the sum of the weights of edges connecting the two sets is maximal. If we associate a variable $x_i\in{\pm1}$ to each of the nodes of the graph, this problem is equivalent to
\begin{align}\label{maxcut}
\max_{\boldsymbol{x}\in\{\pm1\}^N} \sum_{(i,j) \in E} J_{ij}(1- x_i x_j).
\end{align}
Note that the first term in the sum is a problem dependent constant, so the problem is equivalent to the minimization
\begin{align}\label{maxcut2}
    \min_{\boldsymbol{x}\in\{\pm1\}^N}\;C(\boldsymbol{x})=\min_{\boldsymbol{x}\in\{\pm1\}^N}\sum_{(i,j) \in E} J_{ij}x_i x_j.
\end{align}
This minimum coincides with the ground state energy of the Ising Hamiltonian diagonal in the computational basis
\begin{align}\label{maxcutH}
    H = 
        \sum_{(i,j) \in E} J_{ij} \sigma_z^{(i)} \sigma_z^{(j)},
\end{align}
where $\sigma_z^{(i)}$ is the Pauli $z$ operator acting on the $i^{\text{th}}$ qubit. This takes the same form as Eq.~\eqref{costH} with $C(\boldsymbol{x})$ given by Eq.~\eqref{maxcut2}. 

\begin{figure}
    \centering
    \includegraphics[width = 1\columnwidth]{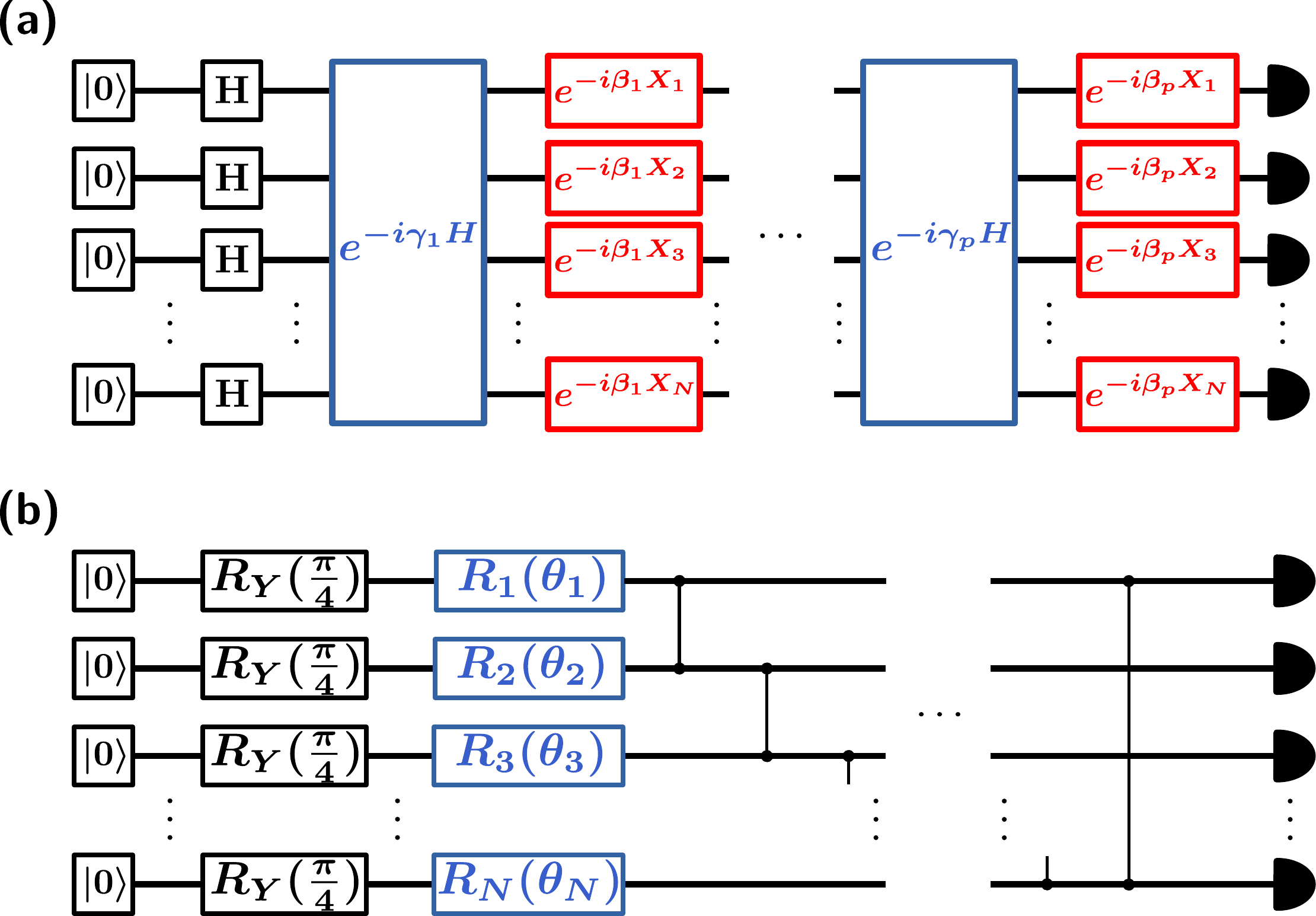}
    \caption{(a) Problem-inspired quantum circuit consisting of a set of Hadamard gates, represented by \textbf{H}, that prepare the $\ket{+}^{\otimes N}$ initial state, followed by a QAOA circuit with $X$ mixers consisting of $p$ repetitions of the cost and mixer gates (represented in blue and red, respectively), and parametrised by $(\boldsymbol{\gamma}, \boldsymbol{\beta})$. (b) Problem-agnostic quantum circuit consisting of a set of of $R_Y(\pi/4)$ gates that generate the initial state of the system, followed by a set of single-parametrized rotations along $X$, $Y$ and $Z$ directions, and a set of nearest-neighbour CZ operations.}
    \label{Fig4}
\end{figure}

We consider three instances of the Max-Cut problem:
\begin{description}
    \item[Instance A]: A fully-connected graph with 5 nodes and weights sampled from a gaussian probability distribution with \added{zero} mean, $\mu=0$, and variance $\sigma^2=1$.
    \item[Instance B]: A 5-regular graph with 8 nodes and weights sampled from two gaussian probability distributions with $\mu = \pm 1$ and $\sigma^2 = 0.3$, where each gaussian is chosen randomly.
    \item[Instance C]: A 3-regular graph with 16 nodes and weights sampled from two gaussian distributions with $\mu = \pm 1$ and $\sigma^2 = 1.0$, where each gaussian is chosen randomly.
\end{description}

In Appendix \ref{AppendixA} we also present some results obtained for the \texttt{ESCAPE} algorithm for a 4-regular graph with 8 nodes and weights sampled from two gaussian probability distributions with $\mu = \pm 2$ and $\sigma^2 = 0.5$, where each gaussian is randomly. The data for the generated problems can be found in \cite{MyGithub}.
\subsubsection{Quantum circuit architectures}
We consider two types of VQC architectures: (i) The problem-inspired \cite{Cerezo2020} architecture called QAOA and; (ii) The problem-agnostic architecture studied in \cite{McClean2018}. Both are shown in Fig.~\ref{Fig4}.

\begin{figure}
	\includegraphics[width=1\columnwidth]{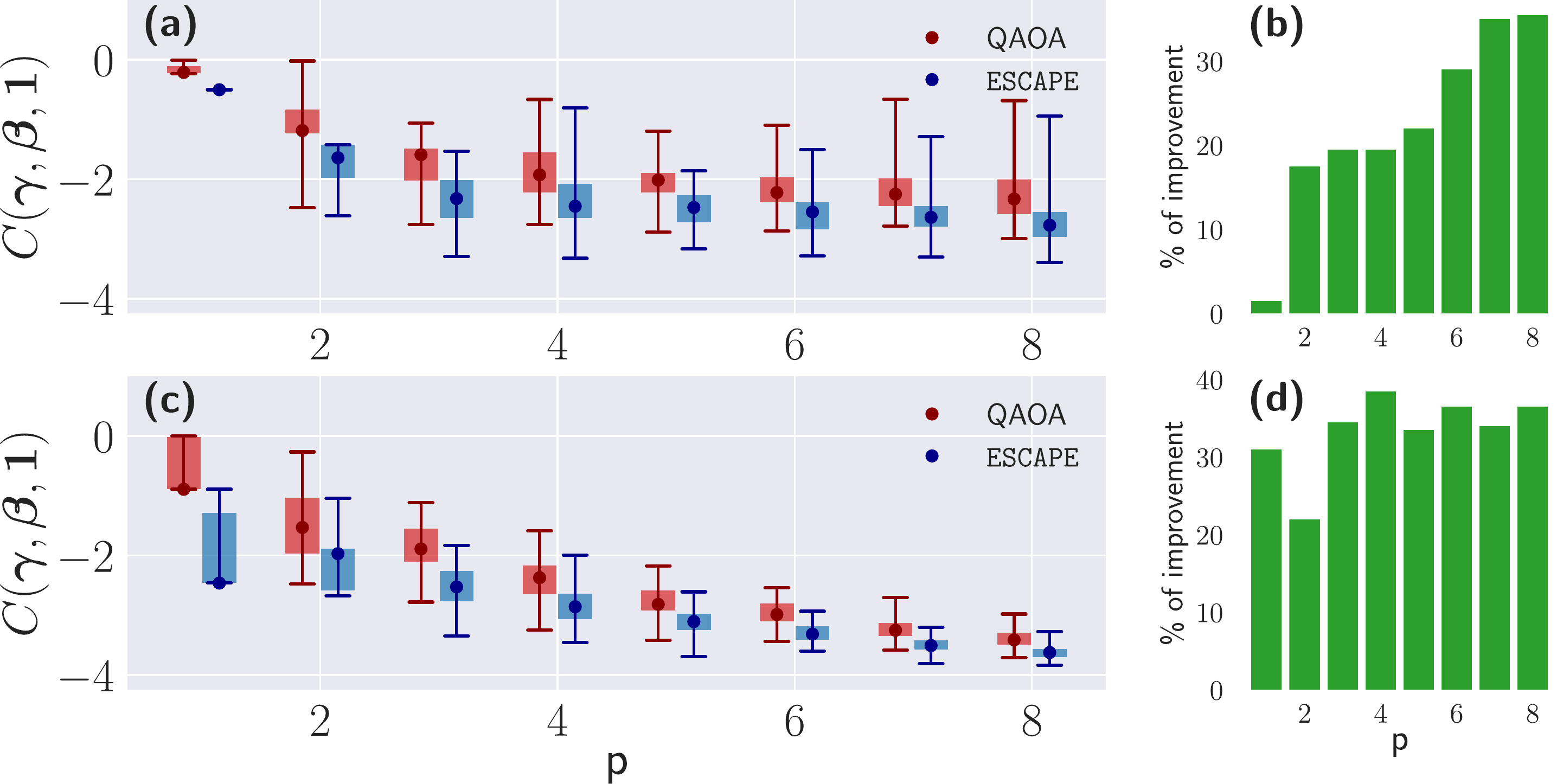}
	\caption{Whisker plots showing the performance of the \texttt{ESCAPE} algorithm using the QAOA circuit architecture with varying depth parameter $p$ for problem \textbf{Instance A}, using standard gradient descent (top, step size $\eta=0.1$) and the Adam (bottom, initial step size $\eta=0.3$). For each $p$, we consider 200 random initializations of the parameters, but only the initializations for which we found improvement are shown. The histograms show the percentage of initializations for which the algorithm successfully escaped the local minimum encountered by the QAOA architecture and converged to a lower cost. The plotted data shows the resulting difference in cost for these initializations only. In both cases, the neural network was optimized for $M=80$ steps in step 3 of the algorithm.}
	\label{Fig5}
\end{figure}

\emph{QAOA architecture}--- The QAOA circuit architecture \cite{Farhi2014} is a popular variational quantum circuit for solving classical optimization problems which corresponds to a quantum circuit ansatz with $2p$ parameters $\boldsymbol{\theta}=(\gamma_1,\beta_1,\cdots,\gamma_p,\beta_p)$:
\begin{equation}
    U(\boldsymbol{\theta})\ket{\psi_0}
        = \prod_{i=1}^p
            U_M(\beta_i)U_C(\gamma_i)\ket{\psi_0}.
\end{equation}
Here the initial state is $\ket{\psi_0}=\ket{+}^N$ and the unitaries are
\begin{equation}
   U_C(\gamma_i) 
        = e^{-i \gamma_i H}
    \ \ \text{and} \ \
    U_M(\beta_i)
        = e^{-i \beta_i H_M},
\end{equation}
with $H$ given by Eq.~\eqref{maxcutH} and $H_M$ the non-commuting mixer Hamiltonian
\begin{equation}
    H_M = \sum^N_{i=1} \sigma_x^{(i)}.
\end{equation}
This choice is standard for unconstrained optimization problems, however other choices of mixer and initial state might be preferable depending on the particular problem one wants to solve~\cite{Hadfield2019, Wang2020}. 

\emph{Problem-agnostic architecture}---For the problem-agnostic architecture, we employed the construction that was used for studying barren-plateaus in \cite{McClean2018}. It consists of a layer of $R_Y(\pi/4)$ gates applied to the product state $\ket{0}^{\otimes N}$ followed by a layer of randomly-drawn single-qubit rotations along either the $X$, $Y$ or $Z$ axes, parametrized by the angles $\boldsymbol{\theta} = (\theta_1, \dots, \theta_N)$. Finally, we apply over each subsequent pair of qubits $i$ and $i+1$ a CZ operation. \\

\begin{figure}
	\includegraphics[width=1\columnwidth]{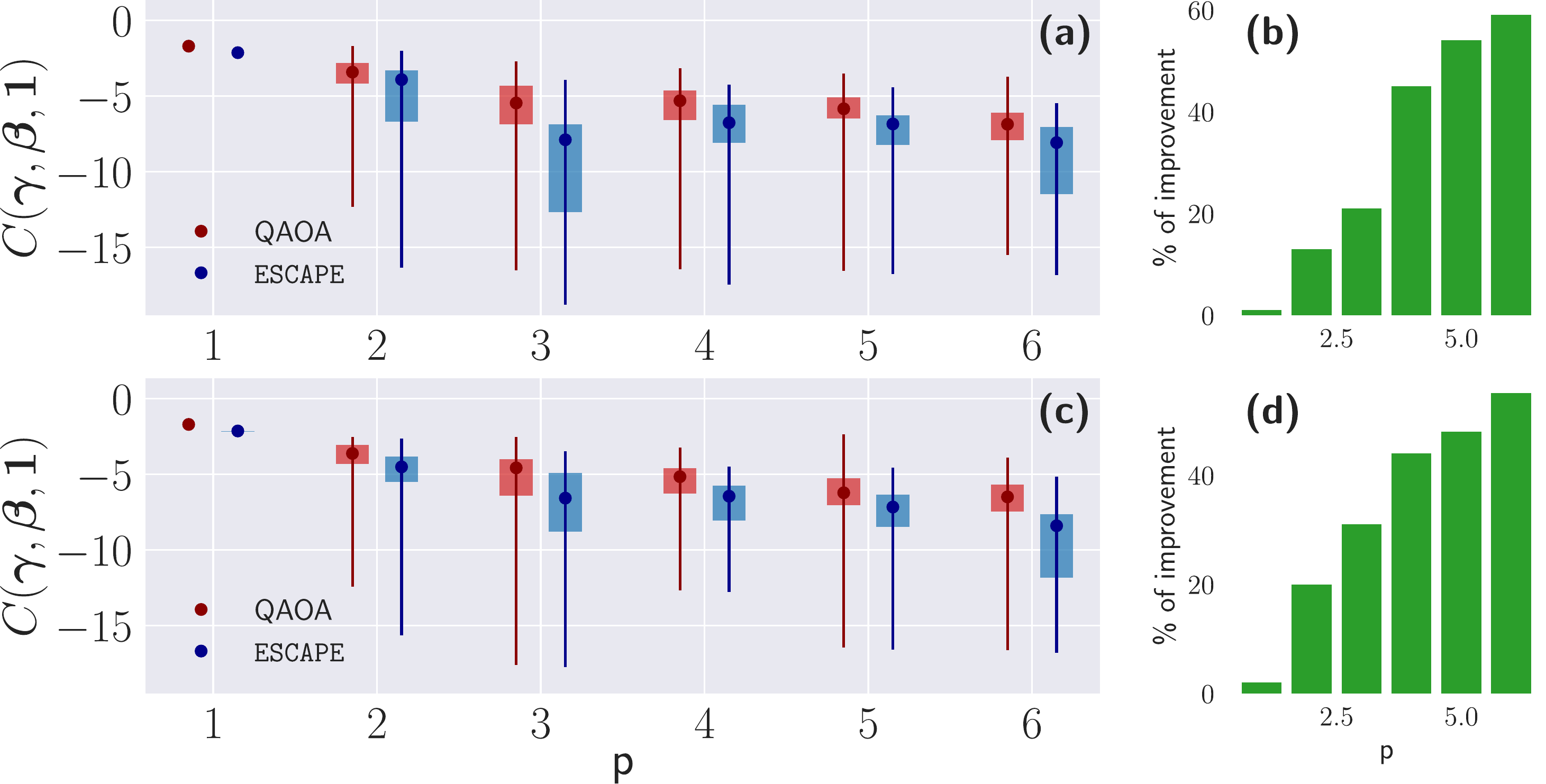}
	\caption{Whisker plot analogous to Fig.~\ref{Fig5} showing the performance of the \texttt{ESCAPE} algorithm for problem \textbf{Instance C}. The optimizer used to obtain these results was Adam with $\eta = 0.1$, and we used (a) $M=50$ and (b) $M=80$ neural network update steps in step 3 of the algorithm. Again, we only show the results for which improvement was found, in agreement with the definition of the \texttt{ESCAPE} algorithm.} 
	\label{Fig6}
\end{figure}

All our calculations have been done numerically in Python using the Pennylane package \cite{Pennylane} for writing and running the quantum circuits, and the corresponding codes can be found in \cite{MyGithub}. More specifically, we have coded both the quantum circuit and the neural network with the TensorFlow library \cite{TensorFlow}, and we have performed the optimization with the Keras package \cite{Keras}. For the 16-qubit architecture we have coded the quantum circuit with NumPy \cite{Numpy} and performed the optimizations with Pennylane based functions.

\subsection{\texttt{ESCAPE} optimization algorithm results}
Here we investigate the performance of the \texttt{ESCAPE} algorithm for both the problem-inspired and problem agnostic architectures defined above. In all our optimizations we use the Heaviside function $g(t) = \Theta(t-x)$, where in the 5- and 8-qubit architectures $x = 150$ (400) in step 4 of the algorithm and $T= 350$, while in the 16-qubit $x=400$ and $T = 800$. For the neural network updates in step 3, we use the standard gradient descent update rule with a step size of $\eta=0.05$.

Figs.~\ref{Fig5} and \ref{Fig6} show the performance of the \texttt{ESCAPE} algorithm for the QAOA architecture for problem \textbf{Instance A} and problem \textbf{Instance C} respectively. For the architecture depths $p$, we ran the \texttt{ESCAPE} algorithm for 200 initializations (100 for the 16-qubit problem) of the parameters, where each parameter is sampled uniformly randomly in $[0,2\pi]$. These figures show the change in energy for those initializations for which the \texttt{ESCAPE} algorithm was able to improve the solution relative to the standard QAOA approach by escaping the local minima encountered in step 2 of the algorithm, and the histograms to the right of each plot show the percentage over the 200 initializations that have been improved. More precisely, our condition for improvement is $\mathcal{C}(\boldsymbol{\theta}_{\text{QAOA}})-\mathcal{C}(\boldsymbol{\theta}^*)>0.1$, where $\boldsymbol{\theta}_\text{QAOA}$ are the parameters of the local minimum of step 2 and $\boldsymbol{\theta}^*$ are the final parameters found by the $\texttt{ESCAPE}$ algorithm, and $\mathcal{C}$ is the cost in the standard QAOA landscape.

\begin{figure}
	\includegraphics[width=1\columnwidth]{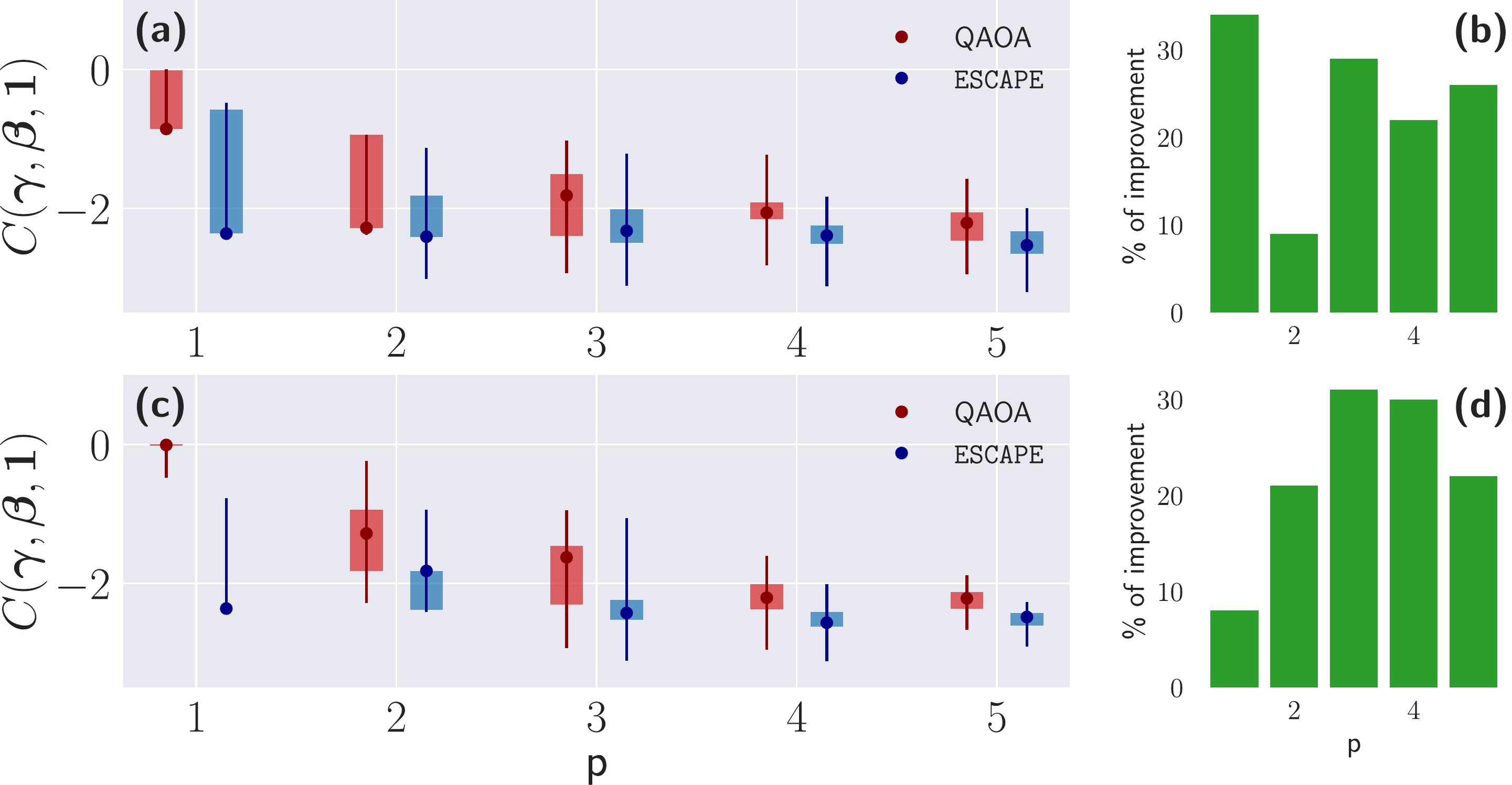}
	\caption{Whisker plot analogous to Fig.~\ref{Fig5} showing the performance of the \texttt{ESCAPE} algorithm for problem \textbf{Instance A} when we add bit flip operations with probability 0.01 after each QAOA architecture iteration. The optimizer used to obtain these results was Adam with $\eta = 0.1$, and we used (a) 50 and (c) 80 neural network update steps in step 3 of the algorithm. Again, we only show the results for which improvement was found such that the percent of improvement associated to each of the neural network updates are shown in (b) and (d).} 
	\label{Noisy}
\end{figure}

\begin{figure}
	\includegraphics[width=1\columnwidth]{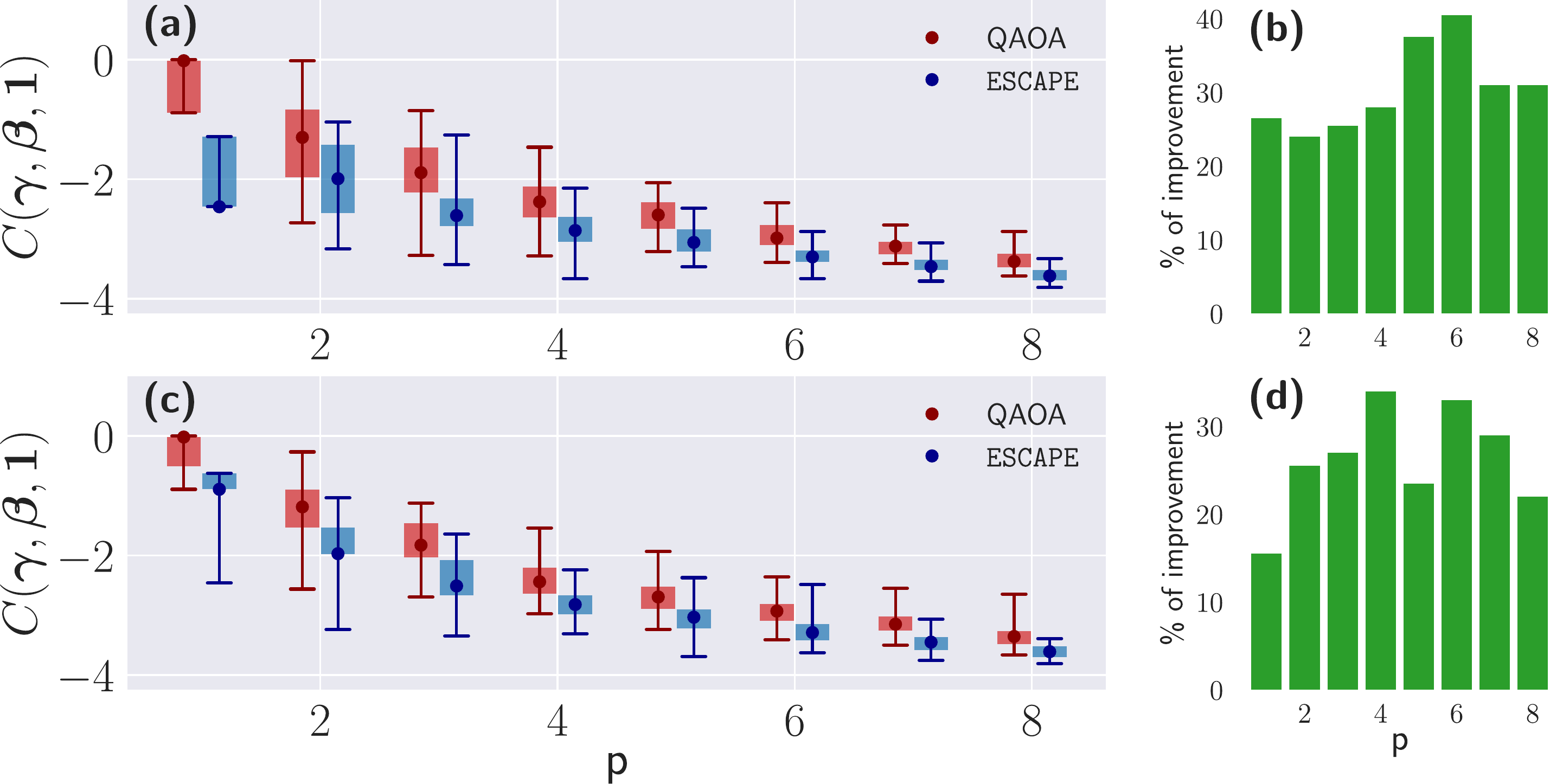}
	\caption{Whisker plots analogous to Fig.~5 showing the performance of vanilla QAOA against the \texttt{ESCAPE} algorithm when using (a) 80 and (c) 150 neural network optimization steps during step 3 of the algorithm, for problem \textbf{Instance A} using the Adam gradient descent subroutine with step size $\eta = 0.1$. In general, both optimization steps perform well, but 80 optimization steps performs better overall, both in terms of the percentage of improvement and in the obtained energy distribution. This highlights that it is important not to over-train the neural network at this step.} 
	\label{Fig7}
\end{figure}

Interestingly, the percentage of local minima that are escaped grows with $p$, which may be due to a greater number of local minima at larger $p$. For the 16-qubit example of problem \textbf{Instance C}, the percentage of improvement can be as large as 60\%, as it can be seen in Fig.~\ref{Fig6}. Moreover, in certain cases, the improvement in the cost can be quite dramatic, especially for low values of $p$. In Fig.~\ref{Fig5} we compare using the standard gradient descent update rule versus the Adam update rule in step 2 of the algorithm. The use of Adam, which combines an adaptive step size and momentum into the gradient descent, results in significantly lower cost values than standard gradient descent, and the \texttt{ESCAPE} algorithm is able to improve these further. We have also noted that it is important not to over-train the neural network in step 3 of the algorithm. This is shown for problem \textbf{Instance A} in Fig.~\ref{Fig7}.

To finalize with the QAOA architecture, we look at the performance of \texttt{ESCAPE} in the presence of noise. For that purpose, we considered problem \textbf{Instance A} while adding noise in the circuit in the form of bit flip operations applied to each qubit independently after each QAOA circuit iteration. More particularly we set the probability of having a bit flip operation in each qubit to 0.01. In Figs.~\ref{Noisy} (a) and (c) we present the results obtained for 50 and 80 neural network optimization steps, respectively. Note that in this case, due to the presence of noise, the energy obtained are in general lower than those shown in Fig.~\ref{Fig5} and moreover the enhancement obtained when increasing the circuit depth diminishes as bigger circuits will have associated more noise. Nevertheless, in this scenario \texttt{ESCAPE} successfully reaches better solutions once the standard approach is stuck in a local minima. In Appendix \ref{AppendixB} we show its performance for a slightly bigger value of noise

\begin{figure}
	\includegraphics[width=1\columnwidth]{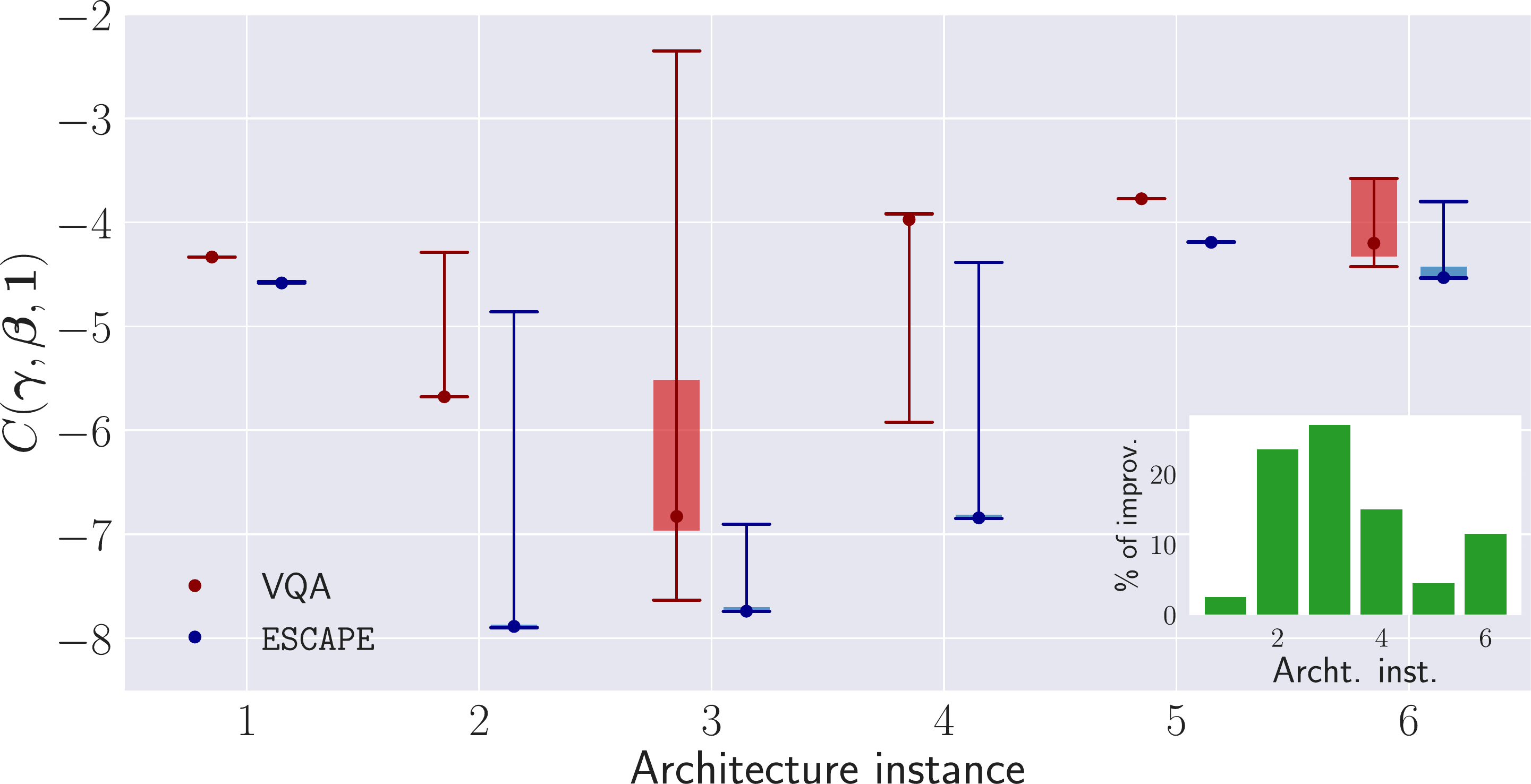}
	\caption{Whisker plot showing the performance of vanilla VQA (in red) against the \texttt{ESCAPE} algorithm (in blue), using the problem-agnostic architecture for the 8-qubit Max-Cut problem \textbf{Instance B}. Each pair of whisker plots corresponds to a different random choice of the rotation gates in the second layer of the architecture. We used 50 neural network optimization and employed Adam with $\eta = 0.1$ for the optimization.}
	\label{Fig8}
\end{figure}

In Fig.~\ref{Fig8} we plot analogous results for the problem \textbf{Instance B} using the problem agnostic architecture. The different whisker plots correspond to different choices of the parameterized rotation gates (either $X$, $Y$ or $Z$ rotations) used in the second layer of the architecture, which are chosen randomly for each whisker plot. The ability of the algorithm to escape the encountered local minima appears to depend strongly on this choice, which we imagine is due to some choices resulting in a badly suited landscape for this particular problem. Nevertheless, for some choices, notably for architecture instances 2 - 4, very large improvements to the cost can be achieved, which can be seen from the large changes in the median cost (architecture instances 2 and 4) and the elimination of the very bad local minima in architecture instance 3. Note that in most of the improvements the median and the interquartile range coincide. This is a consequence of the fact that the generated energy landscapes have a small number of local minima, and the algorithm is taking them most of the time to those ones that we can probably identify as the global minimum.

\subsection{\texttt{GUIDE} optimization algorithm results}
\begin{figure}
	\includegraphics[width=1\columnwidth]{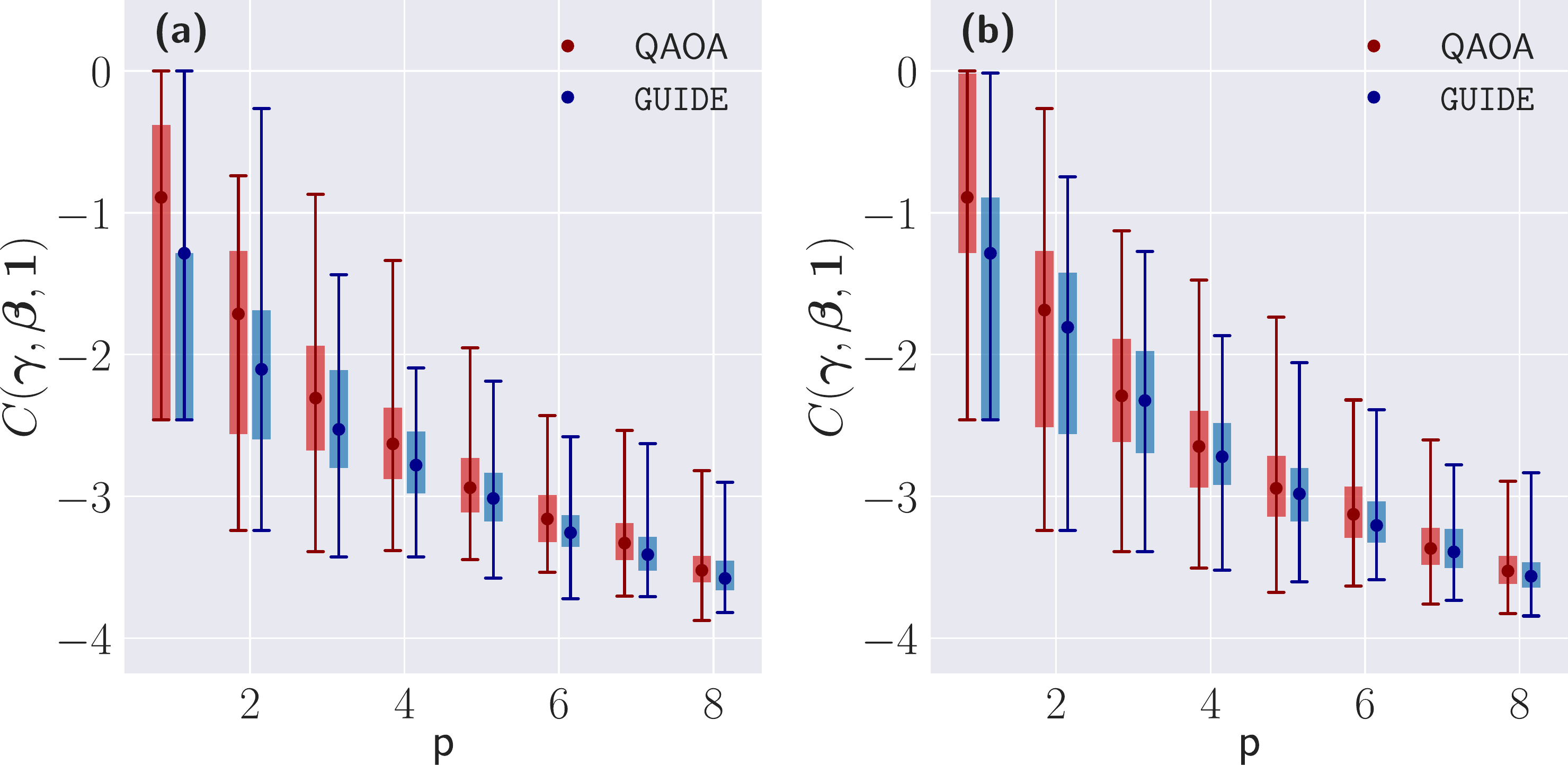}
	\caption{Whisker plot showing the performance of the standard QAOA architecture against the \texttt{GUIDE} algorithm in problem \textbf{Instance A} when using different values of the penalty $\alpha$ during step 2 of the algorithm, namely (a) 0.1 and (b) 0.3. We use Adam (initial step size $\eta=0.1$) and standard gradient descent (step size $\eta=0.05)$ to update the parameters in steps 2 and 3 of the algorithm respectively. In this case, we show the results coming from all the 200 initializations.}
	\label{Fig9}
\end{figure}

Here we investigate the \texttt{GUIDE} algorithm on the 5-qubit problem \textbf{Instance A} using the QAOA architecture. We again consider 200 random initializations of the parameters, sampled uniformly in $[0,2\pi]$. We run both the standard QAOA architecture and the \texttt{GUIDE} algorithm from each of these initilizations and plot in Fig.~\ref{Fig9} the obtained cost after convergence, considering two different penalty terms $\alpha=0.1$ and $\alpha=0.3$. In both cases, the \texttt{GUIDE} algorithm achieves a lower median cost over the 200 initializations and the modification of the cost landscape during the optimization therefore leads to better solutions in general. For some values of $p$, the algorithm fails to find a lower cost than the standard QAOA architecture, although it is possible that this could be mitigated with a better choice of hyperparameters, or a different choice of norm for the penalty term in the cost function.

We remark that in Fig.~\ref{Fig9}, we show the results coming from all the 200 initial random initializations, unlike to the what we did for the \texttt{ESCAPE} results where we only show the improved ones. And the reason for this is that the latter algorithm is used in a similar way to basin-hopping like algorithms, i.e., as an extension to a previous VQA algorithm such that for each initialization we can naturally save the energy coming from the VQA, and directly compare it to the one obtained after executing \texttt{ESCAPE}. This is not possible for \texttt{GUIDE}, as the VQC parameter update is continuously being mixed with the neural network optimization, and therefore to plot the results we have to run separately a neural network-free VQC and \texttt{GUIDE} with the same initialization.

\begin{figure}
	\includegraphics[width=1\columnwidth]{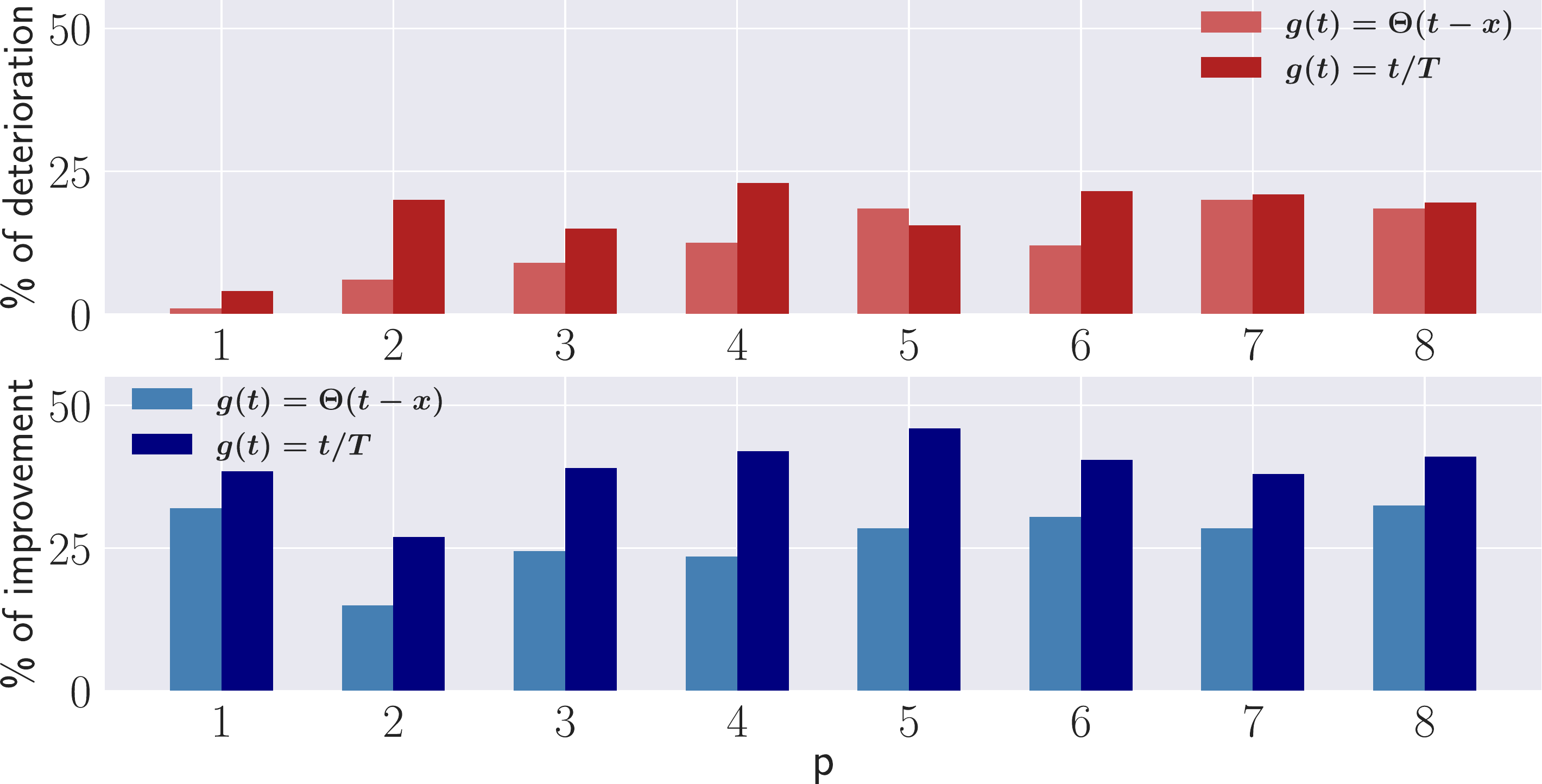}
	\caption{Histogram showing (a) the percent of deterioration and (b) the percent of improvement when using \texttt{GUIDE} ($\alpha = 0.3$) with two different definitions of the $g(t)$ function, in particular a Heaviside function and a linear function, for problems \textbf{Instance A} using Adam gradient descent subroutine ($\eta = 0.1$). In this case, by using the linear function we are able to enhance the percent of improvement but at a price, as the percent of deterioration increases as well. In these plots $T = 350$ and $x= 150$.}
	\label{Fig10}
\end{figure}

Furthermore, we can also modify the definition of the $g(t)$ function and, instead of using a Heaviside function as we have been doing until now, we can use a linear function, i.e. $g(t) = t/T$. While for the case with $\alpha = 0.1$ there is not much improvement in comparison with the Heaviside, for $\alpha = 0.3$ there is a considerable difference. In Fig.~\ref{Fig10} we plot both the percent of improvement (results that have been improved using \texttt{GUIDE} in comparison to QAOA) and the percent of deterioration (results that have been deteriorated using \texttt{GUIDE} in comparison to QAOA) for problem \textbf{Instance A} and $\alpha = 0.3$. This figure suggests that, for this problem, the linear function is more suitable for enhancing the performance of \texttt{GUIDE} as the percent of improvement is higher than the one obtained with the Heaviside function. However, such improvement comes at a price since we also increase the probability of falling into a worse minimum. Nevertheless, the percentage associated to this latter situation is always lower than the percent of improvement.

\begin{figure}
	\includegraphics[width=1\columnwidth]{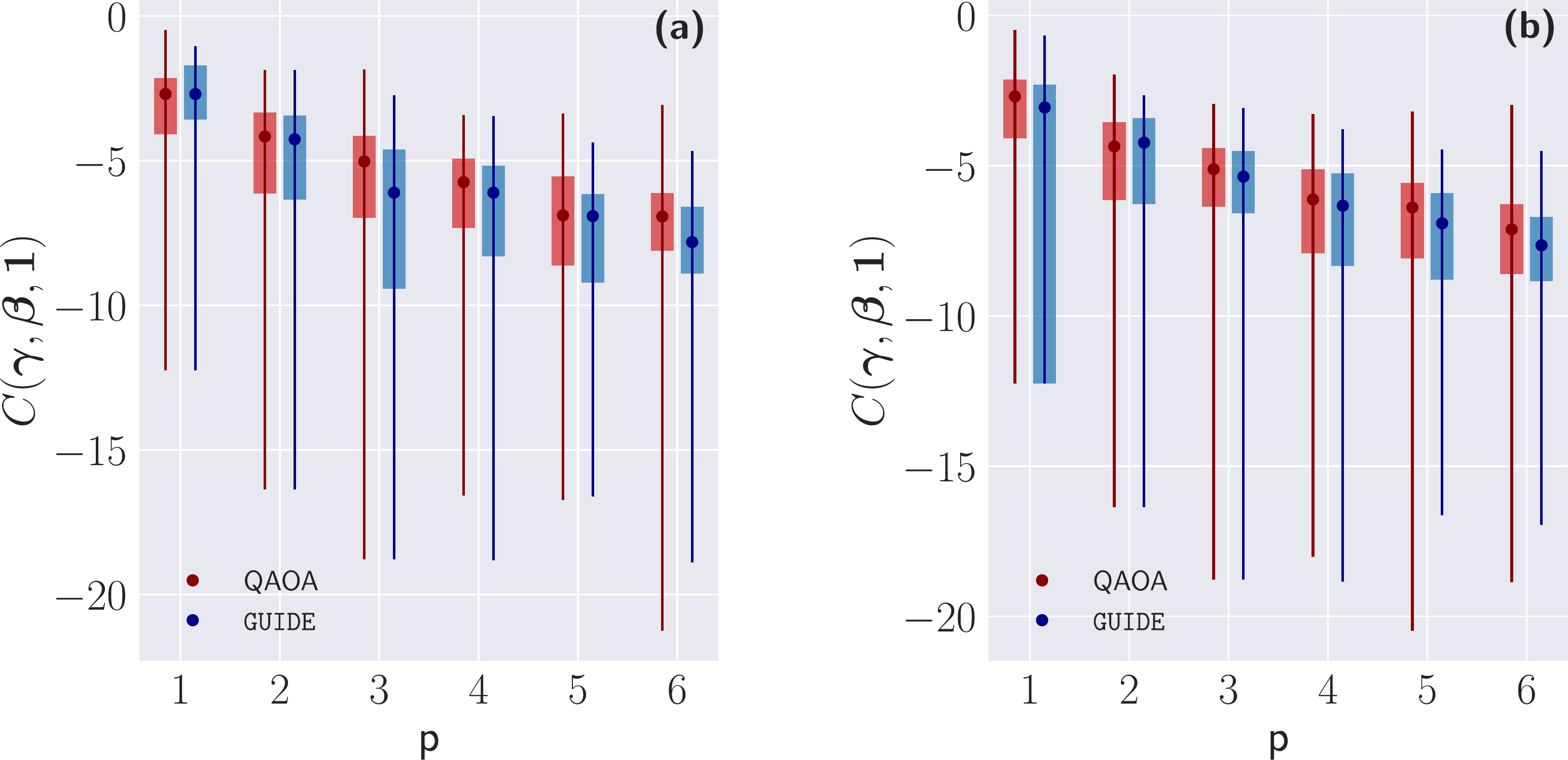}
	\caption{Whisker plot showing the performance of the standard QAOA architecture against the \texttt{GUIDE} algorithm in problem \textbf{Instance C} when using different values of the penalty $\alpha$ during step 2 of the algorithm, namely (a) 0.05 and (b) 0.2. We use Adam (initial step size $\eta=0.1$) and standard gradient descent (step size $\eta=0.05)$ to update the parameters in steps 2 and 3 of the algorithm respectively. In this case, we show the results coming from all the 100 initializations.}
	\label{16qubitGUIDE}
\end{figure}

Finally, in Fig.~\ref{16qubitGUIDE} we present how the \texttt{ESCAPE} algorithm behaves for problem \textbf{Instance C} for (a) $\alpha = 0.05$ and (b) $\alpha = 0.2$. Note from this figure that a value of $\alpha$ that performs better at $p$ than $\alpha' \neq \alpha$, might behave worse for another $p' \neq p$. This can be seen here for $p=3$ for which $\alpha = 0.05$ leads to better solutions than $\alpha = 0.2$, while for $p=1$ we get the other way around as the interquartile range is bigger for $\alpha = 0.2$, implying that in average this value of $\alpha$ leads to more suitable energy landscapes. Thus, this suggests that the selection of the $\alpha$ hyperparemeter not only depends on the particular problem that is being solved, but in general on the specific quantum circuit architecture that is used.

\section{DISCUSSION}\label{discussion}
As is typical with heuristic approaches, our method is sensitive to the choice of hyperparameters that define the algorithm, and it is therefore important that they are selected well. For the \texttt{ESCAPE} algorithm, the main issue encountered with respect to hyperparameter selection is related to over-training of the neural network. More precisely, for the small problem instances that we consider, the neural network is typically able to find the global minimum to the problem for an arbitrary distribution at its input and in this sense is much stronger than the VQC architecture. Training the neural network for too long therefore results in the neural network solving the problem itself, typically leading to a flat (with respect to $\theta$) optimization landscape which prevents optimization of the VQC in the following step. For this reason it is important to limit the number of neural network updates when modifying the landscape, as suggested by Fig.~\ref{Fig7}. For problems with a large number of qubits for which the neural network will not be able to easily solve the problem itself, this may not be a problem however.

In this work we have considered problems that are defined by a single Hamiltonian that are diagonal in the computation basis. The method could however be easily extended to Hamiltonians that are a sum of local Hamiltonians $H=\sum_i H_i$. In such algorithms, the VQC can be used to estimate the cost or gradient for each $H_i$ separately, and  the results are summed to obtain the total cost or gradient. By adding a neural network to the output of the VQC as we have done in this work, one can thus modify the cost landscape by modifying each of the $H_i$ accordingly. Similarly, the gradient with respect to the neural network weights can then be obtained by estimating the gradient for each $H_i$ and summing the results. 

There are a number of ways that one could extend or modify our approach. One interesting possibility is to consider a different quantity to the cost $\mathcal{C}(\boldsymbol{\theta},W)$ when updating the neural network parameters, since simply lowering the cost at that point may result in a deepening of the local minimum. For example, one possibility is to develop methods to update the neural network parameters such that the size of the gradient $\vert\nabla_{\boldsymbol{\theta}}\mathcal{C}(\boldsymbol{\theta},W)\vert$ is increased, thus avoiding reaching areas with small gradients. 

It would also be interesting to consider the effect of using different neural network architectures, since in this work we have considered a very simple model for clarity. One possibility is to use multi-layered feedforward neural networks, although we expect that one would need to go to significantly larger problem sizes to see a considerable difference with the approach presented here. Another possibility is to investigate recurrent neural networks, which could process sequences of quantum circuit outputs at a time, rather than one-by-one as is the case in our approach. It might be also interesting to study the effect of more specialised gradient based methods, such as the Quantum Natural Gradient Descent \cite{Stokes2020}, which are more suitable for the optimization of VQC architectures. 

Finally, we note that it may also be beneficial to pre-process the VQC outputs before inputting them into the neural network, since using real numbers as inputs rather than discrete values as we have done, may result in a smoother training of the neural network. We have not explored this possibility in detail, but consider this and all of the above interesting directions for future work.

\section{CONCLUSIONS}\label{Conclusions}
In this article we have presented two algorithms for improving the overall performance of variational quantum algorithms. These algorithms are based on a trainable post-processing of the quantum circuit outputs with a feedforward neural network, which has the effect of modifying the cost landscape with respect to the quantum circuit parameters. Here, the neural network can be seen as a helper to the variational quantum algorithm, rather than the problem solver itself, since the end goal is to find good quality solutions in the unperturbed cost landscape. 

For small instances of the Max-Cut optimization problem that we have used as benchmarks, our method finds better values of the quantum circuit parameters than the standard approach. Furthermore, since the addition of the neural network carries a negligible quantum circuit sampling cost relative to the calculation of circuit gradients, we expect our algorithms to be a useful tools for improving the optimization of large-scale parameterised quantum circuits. We believe that our general framework is a promising route to overcome some of the difficulties encountered in variational quantum circuit optimization, which we hope inspires future work. 

\section{ACKNOWLEDGEMENTS}
We would like to thank the Pennylane Developers for helping solving the issues regaring the numerical implementation. This work was supported by the ERC AdG CERQUTE, Spanish MINECO (FIS2020-TRANQI, Severo Ochoa CEX2019-000910-S and Retos Quspin), the ICFO-Quside Joint Laboratory in Quantum Processing, the Generalitat de Catalunya (CERCA Program, SGR 1381 and QuantumCAT), Fundacio Privada Cellex and Fundacio Mir-Puig. J.R-D has received funding from the Secretaria d'Universitats i Recerca del Departament d'Empresa i Coneixement de la Generalitat de Catalunya, as well as the European Social Fund (L'FSE inverteix en el teu futur)—FEDER. PH has received funding for this project from the European Unions Horizon 2020 research and innovation programme under the Marie Skodowska-Curie grant agreement No 665884. JB acknowledges the AXA Chair in Quantum Information Science. 

\bibliography{Paper.bib}{}

\begin{appendix}
\section{Performance of \texttt{ESCAPE} in a 8-qubit instance}\label{AppendixA}
\begin{figure}
	\includegraphics[width=1\columnwidth]{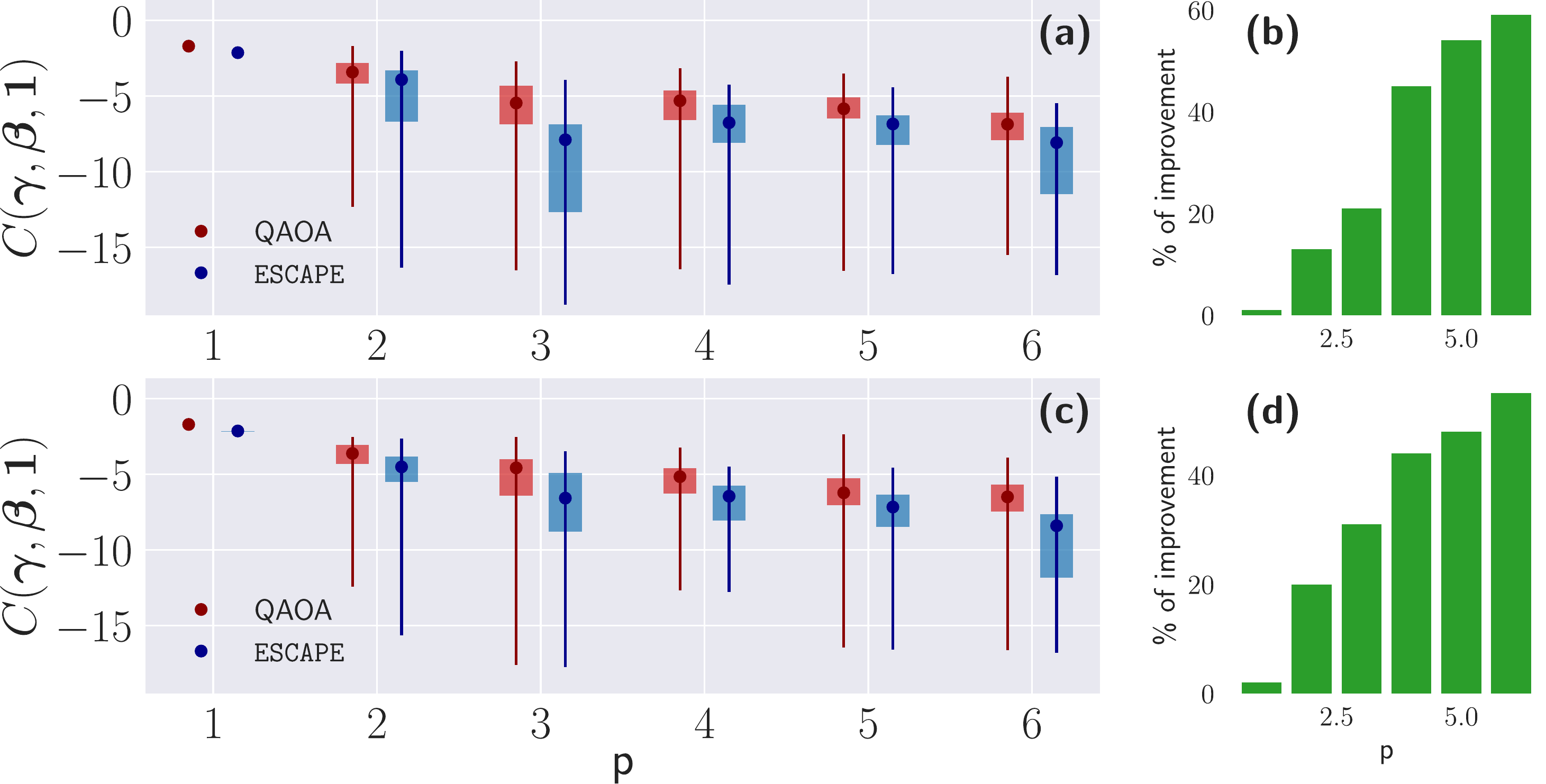}
	\caption{Whisker plot analogous to Fig.~\ref{Fig5} showing the performance of the \texttt{ESCAPE} algorithm for the problem described above. The optimizer used to obtain these results was Adam with $\eta = 0.1$, and we used $M= 25$ neural network update steps in step 3 of the algorithm. Again, we only show the results for which improvement was found, in agreement with the definition of the \texttt{ESCAPE} algorithm.}
	\label{FigAppA}
\end{figure}

Here we present the results obtained for the \texttt{ESCAPE} algorithm for a 4-regular graph with 8 nodes and weights sampled from two different probability distributions with $\mu = \pm 2$ and $\sigma^2 = 0.5$, where each gaussian is chosen randomly. In Fig.~\ref{FigAppA} we show, as customary, those instances for which the \texttt{ESCAPE} algorithm improves the results obtained from direct optimization of the QAOA architecture. Note that in this case, our algorithm allow us to obtain percent of improvements that can be as large as 50\%. 

It is important to remark that the statistics we do in order to perform the Whisker plots involve all the improved results from the initial 200 initializations. This has as main consequence that the results do not show a huge improvement in the convergence when moving from $p$ to $p+1$. In order to see such an enhancement, one should look for the best solution over all the considered initializations per each value of $p$. However, this is not the scope of the present article, where we want to show how in general the overparametrization (introduced by means of the neural network that postprocess the quantum circuit measurements) may be beneficial for escaping local minima.

\section{Performance of \texttt{ESCAPE} with a noisy circuit with bit flip operations applied with probability 0.05}\label{AppendixB}
\begin{figure}
	\includegraphics[width=1\columnwidth]{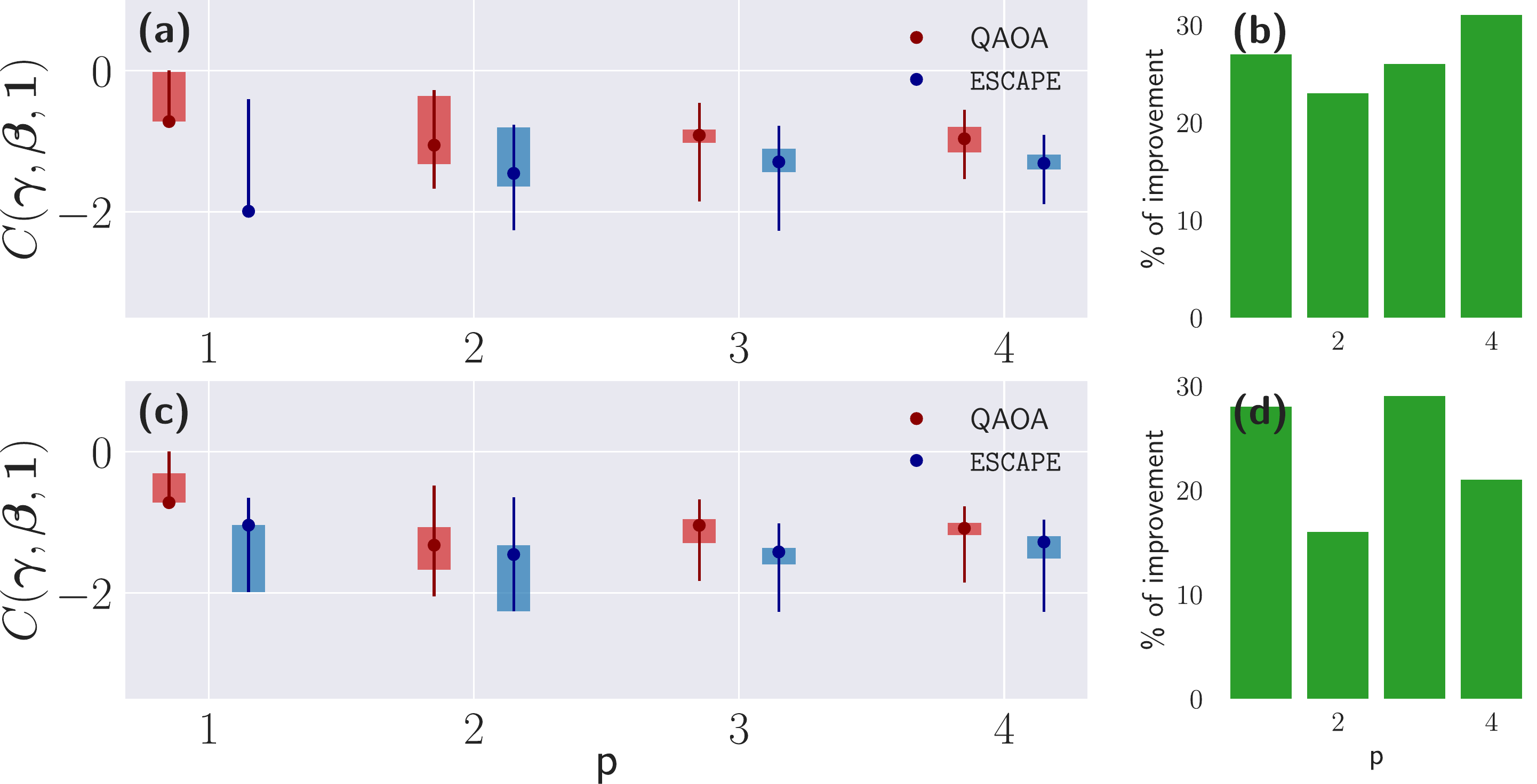}
	\caption{Whisker plot analogous to Fig.~\ref{Noisy} showing the performance of the \texttt{ESCAPE} algorithm for problem \textbf{Instance A} when we add bit flip operations with probability 0.05 after each QAOA architecture iteration. The optimizer used to obtain these results was Adam with $\eta = 0.1$, and we used (a) 50 and (c) 80 neural network update steps in step 3 of the algorithm. Again, we only show the results for which improvement was found such that the percent of improvement associated to each of the neural network updates are shown in (b) and (d).}
	\label{FigAppB}
\end{figure}

Here we present the results obtained for the \texttt{ESCAPE} under the same situation as the one described for Fig.~\ref{Noisy}, but where the bit flip operations are applied on each qubit with probability $0.05$. In Fig.~\ref{FigAppB} we see that the overparametrization introduced with the neural network still allow us to reach lower minima. However, as we increase $p$ the convergence of the method gets reduced as the total noise increases with the circuit depth as well. 

Note that our method coexists with the noise and does not provide a way of avoiding it. The only difference now is that the energy landscape over which we are optimizing now is the one defined by the noise-free quantum circuit and the applied bit flip operations. Thus, for small circuit depths the performance is very alike to the one obtained in Fig.~\ref{Fig5} as small values of noise will not considerably affect the energy landscape defined by the mentioned tandem.
\end{appendix}
\end{document}